\documentclass[preprint,showpacs,preprintnumbers,amsmath,amssymb,floats]{revtex4}
\usepackage{graphicx}
\usepackage[english]{babel}
\usepackage{amsmath}
\usepackage{amssymb}
\usepackage{color}

\newcommand{\fk}{\mathbf{k}}
\newcommand{\fq}{\mathbf{q}}

\newcommand{\be}{\begin{eqnarray}}
\newcommand{\ee}{\end{eqnarray}}

\newcommand{\om}{\omega}
\newcommand{\ek}{\epsilon_{{\bf k}}}
\begin{document}

\title{Pseudogap in Cuprates in the Loop-Current Ordered State}
\author{C. M. Varma}
\affiliation{Department of Physics, University of California, Riverside, CA.}
\date{\today}

\begin{abstract}
Scanning tunneling microscopy (STM) has revealed that the magnitude of the {\it pseudo-gap} in under-doped cuprates varies spatially and is correlated with disorder. The loop-current order, characterized by the anapole vector ${\bf \Omega}$, discovered in under-doped cuprates  occurs in the same region of the temperature and doping as the pseudo gap observed in STM and ARPES experiments. Since translational symmetry remains unchanged in the pure limit, no gap occurs at the chemical potential. On the other hand  for disorder coupling linearly to the different possible orientations of ${\bf \Omega}$, there can only be a finite temperature dependent static correlation length for the loop-current state at any temperature. This leads to formation of domains of the ordered state with different orientation and magnitude of ${\bf \Omega}$ in each. For the characteristic size of the domains much larger than the Fermi-vectors $k_F$, the boundary of the domains leads to forward scattering of the Fermions. Such forward scattering is shown to push states near the chemical potential to energies both above and below it leading to a {\it pseudo-gap} with an angular dependence which is maximum in the $\pm|\hat{x} \pm \hat{y}| =0$ directions because the single-particle energies are degenerate in these directions for all domains. The magnitude of the average gap systematically increases with the square of the average loop order parameter measured by polarized neutron scattering. This result is tested. A unique result of the gap due to forward scattering is the lack of a bump in the density of states at the states at the "edge" of the pseudo-gap so that the depletion of states near the chemical potential is recovered only in integration up to the edge of the band. This is also in agreement with a variety of experiments. Some predictions for further experiments are provided. Due to the finite correlation length, low frequency excitations are expected at long wavelength at all temperatures in the "ordered" phase. Such fluctuations motionally average over the shifts in frequencies of local probes such as NMR and muon resonance expected for a truly static order. 
\end{abstract}
\maketitle
\section{Introduction}
An Ising model with disorder which couples linearly to the order parameter has no long-range order down to zero temperature for arbitrarily small disorder in 2 or lower dimensions \cite{Imry-Ma}. Instead domains of order in different orientations form. This argument generalizes to discrete models with more possible equivalent directions.  A transverse field discrete model \cite{senthil} also does not change the essential results. These works are concerned with insulators. The new question that is posed here is: suppose there is a metal which in the pure limit has a phase transition of its collective degrees of freedom with lattice symmetry preserved, so that in the low temperature phase the single-particle spectra has a zero derivative of the density of states at the chemical potential. Suppose now that the metal has quenched disorder which couples linearly to the order parameter so that the low temperature states has finite size domains with order oriented in different directions . What then is the spectra of one-particle excitations in the disordered problem?

In such a disordered situation, there is a large density of static fluctuations of the order parameter below the ordering temperature of the pure problem peaked around momenta ${\bf q} = 0$ with a width related to the size of the domains. I show below that a reduction in the spectral weight of one-particle spectra of the fermion occurs at the chemical potential provided there is finite coupling to in the ${\bf q} \to 0$ limit for fermions scattering off such fluctuations. This is the only way to get around Bloch's or Floquet's theorem for electronic structure in periodic systems that gaps (other than those due to superconductivity or due to some unknown state which also generates new quantum numbers) occur only at Brillouin-zone boundaries, not at arbitrary chemical potential. The idea is seen to be reasonable from the well known result \cite{Ashcroft-Mermin}  that the exchange self-energy due to unscreened Coulomb interaction ($\propto 1/q^2$) leads to the density of states near the chemical potential $N(E) \to 0$ as $\log^{-1} |E-E_f|$  in $d=3$ and as $1/|E-E_f|$ in $d=2$. 

This basic idea is applied to the "pseudo-gap " phenomena observed in under-doped cuprates. A large anisotropic  reduction  in the single-particle spectral weight measured in Angle-Resolved Photoemission (ARPES) experiments \cite{arpes1},\cite{arpes2} at the chemical potential, over a wide range of doping $x$, occurs in all cuprates starting at temperatures $T$ below about $T^*(x)$, below which they also show changes in transport and thermodynamic properties from those above \cite{timusk-rev}, \cite{tallon}, \cite{norman}.  This is one of the many astonishing features of the universal phenomenology of the cuprates. 

The pseudo-gap was first discovered through Knight-shift or magnetic susceptibility experiments \cite{alloul} and confirmed by Scanning tunneling \cite{stm-fischer, stm-rev, stm-hoffman} and ARPES \cite{arpes1},\cite{arpes2}spectroscopy . It has been the 
{\it siren-song} in the high $T_c$ problem. Since tunneling and ARPES experiments have revealed it to be a reduction in the single-particle density of states, the idea that the pseudo-gap is some kind of "spin-gap", meaning a reduction in the magnetic susceptibility through particle-hole correlations, as in the Resonant Valence Bond type of ideas \cite{lee-rmp} is untenable. The anisotropy of the reduction in the single-particle density of states can in turn easily seduce one to other untenable directions. The reduction is maximum close the $(\pi,0)$ and equivalent directions and minimum in the $(\pm \pi/4, \pm \pi/4)$ directions. This is what would be expected for a spin-density wave order of the same (or nearly the same) ${\bf Q}$ vectors as the insulating AFM phase of the cuprates. This is indeed what is found in well defined calculations on the Hubbard model \cite{Tremblay}, \cite{millis}. But there is no sign of AFM order or even large AFM correlation lengths \cite{balatsky-bourges}  at  $T \lesssim T^*(x)$ for which the phenomena is observed. For some cuprates charge density modulations that occur  \cite{tranquada}, \cite{jullien}, \cite{ghirenghelli}  in some range of $x$ at lower temperatures have neither the symmetry nor the amplitude to compare with measured single-particle spectra or with thermodynamic experiments. They also appear to be stronger or only exist when a sufficiently strong magnetic field is applied and therefore not an issue for consideration of the pseudo-gap.
The anisotropy of the pseudo-gap is  also similar to the anisotropy in the density of states of the d-wave superconductivity in the cuprates below temperatures $T_c$, which are over most of the region of $x$ well below $T^*(x)$. So, one might  think that long superconducting correlations of amplitude or phase might be responsible for the pseudo- gap \cite{emery-kivelson, norman-randeria, reber-dessau, englebrecht}. But there is evidence of measurable superconducting correlations only in regions of at most $\pm 20$ K above $T_c(x)$, i.e. far below $T^*(x)$ for most range of $x$ for which pseudogap is observed. 

Based on mean-field calculations on a three-orbital model for cuprates, it was suggested that $T^*(x)$ is actually a line of transitions below which a novel state emerges and that this line terminates at a quantum-critical point within the dome of superconductivity \cite{cmv-simon, simon, cmv-prb2006}. On variants of the basic model, such states have recently been shown to be the ground state in similar range of doping  through variational Monte-Carlo calculations in asymptotically large lattices and exact diagonalization in clusters of 24 sites \cite{weber}. Extensive experiments \cite{bourges} \cite{greven-prb} \cite{bourges-review} \cite{kaminski} in a number of families of cuprates have revealed an order consistent with such a proposed state  in under-doped compounds setting in at a temperature consistent with $T^*(x)$. This state breaks time-reversal symmetry through generation of two current loops in each unit-cell of opposite chirality ({\it Anapole Order}), while preserving translational symmetry. Thermodynamic evidence of a change of symmetry at a temperature consistent with $T^*(x)$ has also been presented \cite{leridon} \cite{shekhter}. Signatures of time-reversal breaking accompanying this transition are observed in a number of cuprate families through Kerr effect \cite{kerr}. On one sample of Bi2201 \cite{He-arpes-kerr},
ARPES, Kerr effect and time-resolved reflectivity measurements have all been performed consistent with symmetry breaking and growth of an order below $T^*$. The magnitude of the principal order parameter in this state measured by polarized neutron scattering is large, as much as $0.1 \mu_B$ per loop or a staggered order of $O(0.2) \mu_B$ in each unit-cell \cite{bourges, greven-prb} at low temperatures. At this magnitude the free-energy reduction due to loop order is larger than the measured superconducting condensation energy near optimal doping \cite{cmv-lz}. It is therefore natural to seek to understand the pseudogap phenomena in terms of the properties of fermions in such a state. But since translational symmetry is not broken, no gap at the chemical potential can occur for such a state in the pure limit. In the fluctuation regime of the loop order transition, forward scattering does indeed lead to pseudo-gap behavior \cite{cmv-prb2006} but not in the ordered phase.

There is however one important fact which makes the fluctuation to the loop order state persist for $T \lesssim T^*(x)$. The order parameter of the loop ordered state is a discrete order parameter (with 4 possible directions in a unit-cell) in an extremely anisotropic problem which may be regarded as two dimensional.
For any disorder coupling linearly to discrete order parameter in $d=2$, there is no long-range order even at $T=0$ \cite{Imry-Ma}. In effect, the fluctuations in the vicinity of the phase transition in the pure limit are frozen (with important temperature dependent modifications and slow fluctuations as a function of time) down to $T=0$. It is shown in this paper that in this situation, anisotropic reductions of the density of states must occur for temperatures below the putative loop current order in the pure problem. There is never a true gap at the chemical potential but a change in the form of the spectral function, which is compatible with the experimental results.  The fact that different cuprate compounds and even different samples of the same compound show differences in details of the change in transport and thermodynamics and indeed in the value of $T^*(x)$ suggests that disorder might play a crucial role. In one family of cuprates, Bi$_2$Sr$_2$CaCu$_2$O$_{8+\delta}$ (Bi2212) in which scanning tunneling microscopy (STM)measurements \cite{stm-rev},\cite{stm-hoffman} has been very successful, vivid evidence of disorder and its correlation with the local pseudo-gap has been presented. Various other experimental tools reveal considerable disorder in essentially all the cuprates \cite{disorder}. The importance of defects to the properties of cuprates has been variously emphasized \cite{gorkov}. As discussed below, the nature of the disorder induced in the $CuO_2$ layers deduced by STM measurements is consistent with that which couples linearly to the loop current order parameter.

\begin{figure}
\includegraphics[width=0.5\textwidth]{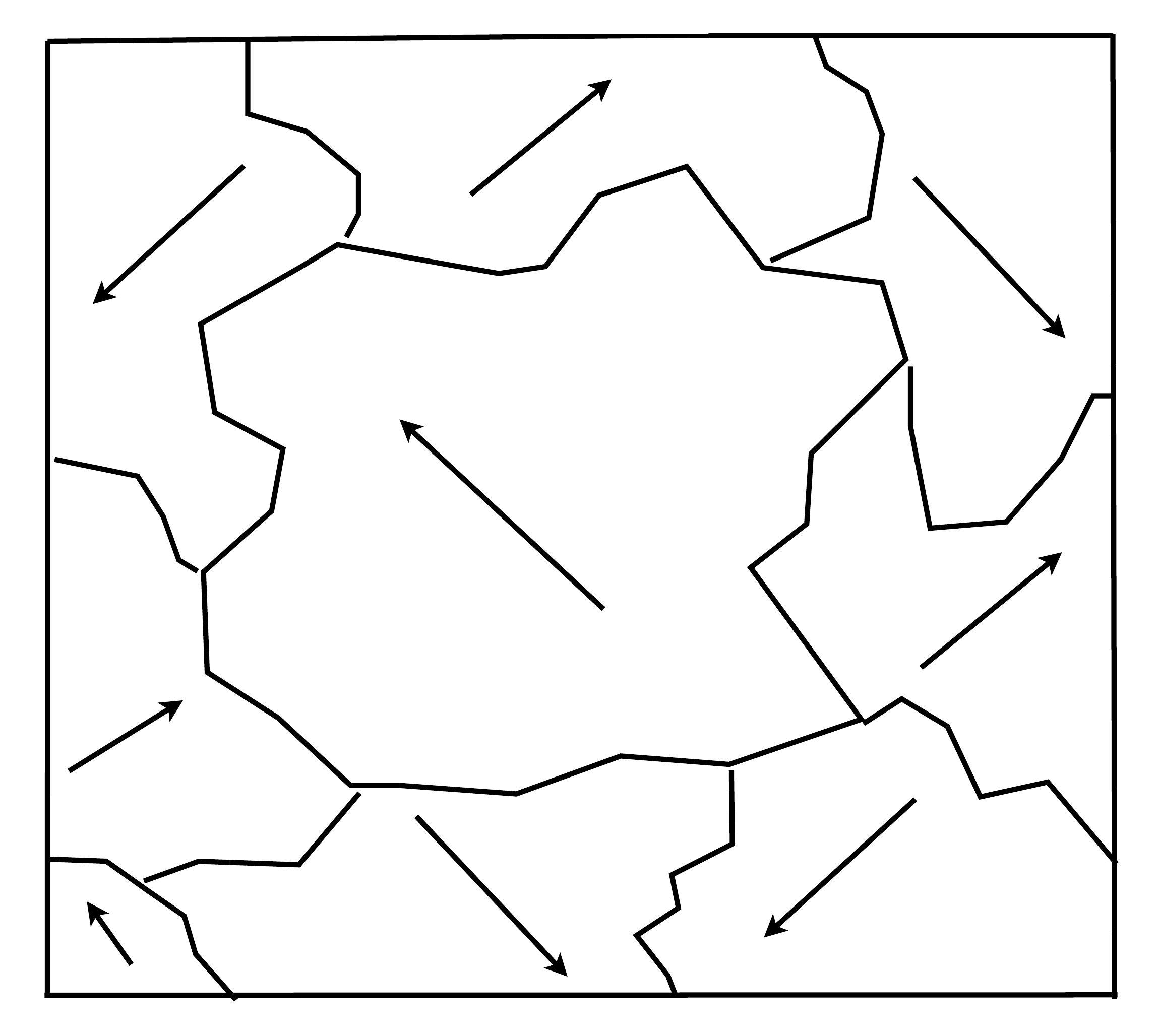}
\caption{The domain pattern for loop current order due to disorder favoring one or the other domain locally. The arrows represent the local order parameter as specified in Fig. (\ref{Order}).}
\label{domains}
\end{figure}

The theoretical problem posed for the single-particle spectra in the presence of local disorder favoring one of the four directions of order is illustrated in Fig.(\ref{domains}). 
Such a problem is of interest more generally than in the cuprates. Domain of a given order is surrounded by domains of a different order by a complicated boundary. The characteristic size of the domains is the correlation length determined by the disorder. The fermions have wave-functions inside a domain which are nearly the eigenstates of the Hamiltonian for that domain but scatter to the eigenstates for the Hamiltonian of the adjoining domain due to {\it static} fluctuations whose correlation function is calculated using the structure factor of the domains and with matrix elements derived from the microscopic Hamiltonian. The global wave-functions are the average over domains of sums of wave-functions in different domains which go from one domain to another continuously, assuming no localization. 

Although neutron scattering presents clear evidence of a change of symmetry in going across the pseudo-gap boundary and thermodynamic signatures of the change in symmetry consistent with it are observed, there is an important class of measurements using local probes such as muon resonance, as well as Nuclear Quadrupole Resonance and Nuclear Magnetic Resonance, which do not observe the magnetic fields expected for long-range static order \cite{muons}, \cite{nqr},\cite{nmr}. The time-scale of these experiments is typically 5 orders of magnitude finer than the resolution of the "elastic" neutron diffraction experiments. It is proposed that the dynamics of the glassy phases such as in Fig. (\ref{domains}) is such that the expected shifts in the resonance lines are motionally narrowed \cite{NMRbook} over alternate directions of the local magnetic field. This can be tested by further experiments. 

This paper is organized as follows. In the next section, I discuss the  symmetry of random fields due to local strains which couple to the loop-current order parameter for the cuprates. I also provide a summary of what is understood about the static structure factor $S(q)$ and the inverse correlation length $\kappa$ of the random field discrete models, as these are necessary inputs to the calculation of the single-particle spectra.  Sec. III presents the principal new theoretical calculations of the paper - that of the single-particle spectral function due to scattering from the domain walls.  In Sec. IV, a comparison of the spectral function and results following from it are compared with ARPES, STM and thermodynamic and transport experiments.  The concluding Sec. V contains a few predictions for further experiments.
\section{Correlations in the Random field Discrete Models}

\subsection{Coupling of Defects to the Loop-
Current Order Parameters}
\begin{figure}
\includegraphics[width=0.7\textwidth]{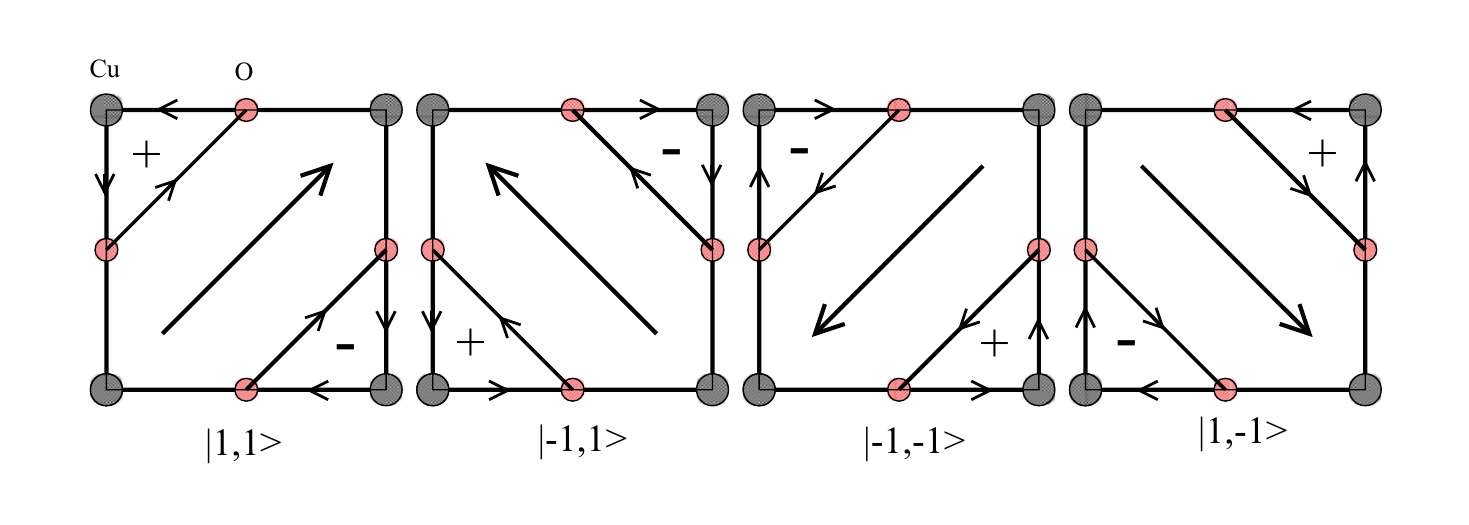}
\caption{The current patters and the direction of the anapole ${\bf \Omega}$ in the four possible domains of order in the loop-current phase. In Fig. (\ref{domains}, the four direction of arrows in the different domains represent these directions.}
\label{Order}
\end{figure}

The observed order parameter in the under-doped cuprates  is specified by the magnitude and direction of the {\it anapole} order, specified in each cell $i$ by the polar-time-reversal odd vector
 \be
 \label{order-para}
 {\bf \Omega}_i = \int_{cell-i} d^2r~ ({\bf L}({\bf r}) \times {\bf \hat{r}}).
 \ee
 Here ${\bf L}({\bf r})$ is the magnetic moment distribution due to two orbital current loops formed between the O-ions and the Cu-ions in each cell; the four possible directions of ${\bf \Omega}_i$ are  shown in Fig. (\ref{Order}). 
The four different possible ${\bf \Omega}$ are time-reversal odd polar vectors pointing in the $\pm \hat{x}', \pm \hat{y}'$-directions. Here $\hat{x}'=(\hat{x}+\hat{y})/\sqrt{2}; \hat{y}'=(\hat{x}-\hat{y})/\sqrt{2}$. 
The {\it linear} coupling of lattice distortions at an impurity cell $i$ to the possible local order parameters in that cell \cite{shekhter} may be seen as follows:  A free-energy invariant with a lattice distortion ${\bf u}_i(B_{2g})$ in cell $i$ of $B_{2g}$ symmetry (which transforms as $xy$, i.e. a monoclinic distortion) is,
\be
\label{dis-Ham1}
\sum_i \gamma ~ {\bf u}_i(B_{2g}) (\Omega^2_{i,x'} - \Omega^2_{i,y'}) 
\end{eqnarray}
Here $\gamma$ is a coupling constant, assumed positive in full generality. Suppose lattice distortion $<{\bf u}_i(B_{2g})>$ is non-zero at a small concentration $c$ of unit-cells with random signs distributed over the defective sites $i$. Then its sign determines which among the two pairs of states $\pm\Omega_{i, x'}$ or  $\pm\Omega_{i,y'}$ is favored locally. In 2 dimensions, for arbitrarily small disorder, domains must form between states of $|\Omega_{x'}| \ne 0 $ and $|\Omega_{y'}| \ne 0$ seeded by the defective sites and of a characteristic size determined by $c$, the disorder strength ${\bf u}_i(B_{2g})$ and the temperature dependent correlation length in the pure problem, as discussed below.  Note that such a time-reversal invariant disorder does not distinguish between $+\Omega_{x'}$ state and $-\Omega_{x'}$ state, or between $+\Omega_{+y'}$ and $-\Omega_{y'}$, which must form with equal probability. The difference from the case of a four state model with disorder favoring one or the other states linearly is that in the latter case, domain of a given state has boundaries to domains of  the three other states, while in the present case, it has boundaries to only two other states; the third "boundary" occurs only at corners in the limit $T \to 0$. At finite temperature entropy will force a finite boundary between such domains also. This issue is not considered in detail as it is not crucial to the principles in the calculation of the single-particle self-energy.

Local monoclinic distortions consistent with $B_{2g}$ symmetry have been identified on the surface through STM measurements \cite{stm-rev}, \cite{stm-hoffman}. The effect of these defects is to change the local energies of the O and Cu orbitals and to change the transfer integral between a given Cu and its O neighbors or between two O neighbors. Direct demonstration of the correlation of the variation in the magnitude of the pseudo-gap and disorder has been obtained in STM measurements \cite{stm-rev}, \cite{stm-hoffman}. A correlation length in the pseudo-gap magnitude of about 20-25 unit-cells is inferred from the real space maps at the surface. These local measurements are not sensitive to the symmetry of the loop-current order parameter, unlike the neutron scattering measurements, which on the other hand average over large volumes of the sample and give no information about the domains except through the correlation length of the order parameter, which varies \cite{bourges-review} from more than about 30 unit-cells for the measured samples in $YBa_2Cu_3O_{6+x}$ and $HgBa_2CuO_{4+x}$ to about 10 unit-cells in the sample measured of Bi(2212) to only about 4 unit-cells in the sample measured of $La_{1.9}Sr_{0.1} CuO_4$.  

The magnitude of the pseudo-gap will depend on the local magnitudes of ${\bf \Omega_i}$ which are determined by the magnitude of disorder. Variation in the magnitude of ${\bf \Omega}_i$ automatically follow due to the variation in the magnitude of ${\bf u}_i$. But the magnitude variations also depend on $|{\bf u}_i|^2$ of various symmetry which couple quadratically to the possible order parameters. The latter do not however necessarily lead to domain formation for arbitrarily small magnitude of disorder, unlike the linear couplings to disorder. We will however concentrate only on the linear couplings, for a simple model of disorder with concentration $c$ of defective sites of equal magnitudes of disorder $\pm<{\bf u}>$ distributed with equal probability.   

\subsection{Static Structure factor of the Random-Field Discrete Models}
As shown in the next section, the self-energy of fermions due to scattering from domain walls can be calculated using the static structure factor of domain walls $S_w(q)$, which can be calculated from the static structure factor $S(q)$ of the domains.  Approximate theory of the static structure factor in a random field problem was given long ago, by Lacouer-Gayet and Toulouse \cite{Toulouse}  and by various others using more sophisticated methods \cite{dedominicis}. A simple version of the calculation for the static structure factor $S(q)$ including the region below the critical point has been given by Halperin and Varma \cite{HV} and by Vilfan and Cowley \cite{cowley}; the latter also provide a self-consistent determination of the inverse correlation length of domains $\kappa$ which depends both on the distribution of disorder and the inverse correlation length of the pure problem. They also give a summary of the experimental evidence for the effects of random fields on antiferromagnets, both in two and three dimensions. The results needed for the present paper are summarized here. 

The static Structure factor has two contributions, $S_e(q)$ directly due to independent disorder fields from the impurity sites, and $S_c(q)$ due to the re-orientation effect of the random fields induced by the impurities:  
 \be
 \label{Sq}
 S(q) &\approx &S_e(q) + S_c(q), \\
 S_e(q) &=& \frac{T}{J} \Big( \frac{<<\Omega>_{T}>^2>_{av}}{q^2 + \kappa^2}\Big), ~~
 S_c(q) =  \Big(\frac{<h^2>}{J^2}\Big) \Big(\frac{<<\Omega>_{T}>^2>_{av}}{q^2 + \kappa^2}\Big)^2 .
 \ee 
 \begin{figure}
\includegraphics[width=0.5\textwidth]{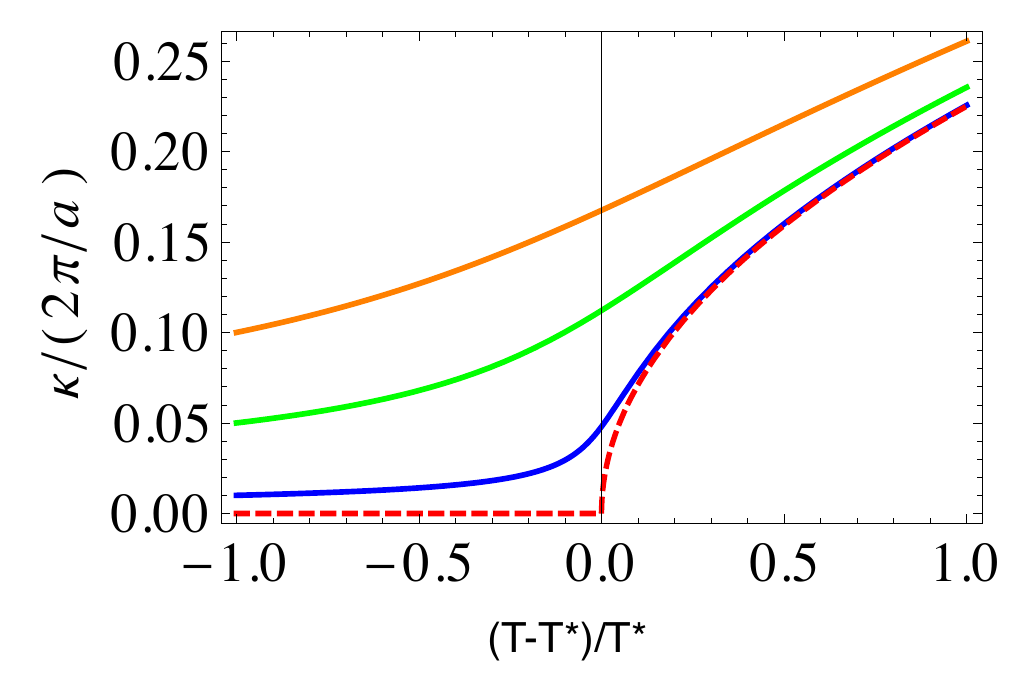}
\caption{The variation of the inverse correlation length $\kappa$ with temperature for various values of $\kappa_0$ given in Eq. (\ref{kappa}). The red-dashed line is for the no disorder case, $\kappa_0 = 0$ at $T_c$ and below, the blue, green and orange are for $\kappa_0/(2\pi/a) = 0.01, 0.05, 0.1$, respectively. The mean-field variations above $T_c$ are also shown.}
\label{fig:kappa}
\end{figure}
$<<\Omega>_{T}>^2>_{av}$ is the average of the mean-square order parameter \cite{footnote}; the first average is over temperature for a fixed disorder and the second is the average over disorder. $J$ is the near neighbor interaction of the pure model,  $<h^2>$ is the dimensionless mean-square random field on the cells defined as
\be
\label{h2}
<h^2> = c  <\big(\gamma <u>{\Omega}_d\big)^2>.
\ee
Here $c$, as defined earlier, is a concentration of defective cells and $<{\Omega}_d^2>$ is the dimensionless mean-square order parameter in the defective unit-cells averaged over the defective cells induced by the lattice distortion $<u>$, as discussed above in relation to Eq. (\ref{dis-Ham1}). It will in general have a magnitude larger than that given by the order parameter in the pure limit.  $\kappa$ is the inverse of the effective correlation-length of domains, measured in units of the reciprocal lattice units $(2\pi/a)$. It gives the characteristic (inverse) size of the domains.  $\kappa(T)$ at low temperatures may be identified as the inverse correlation length of the order parameter measured by neutron scattering which as mentioned earlier is similar to that measured for by STM for Bi(2212). 

Vilfan and Cowley have derived, through a simple self-consistency argument and a mean-field approximation, that for  $d=2, ~\kappa$ is given by 
\be
\label{kappa}
t \kappa^2 +\kappa_0^2 &=& 2 \pi^2 \kappa^2(\kappa^2 -\kappa_0^2), ~~
\kappa^2_0 = \frac{1}{4 \pi^{3}} \frac{<h^2>}{T^{*2}} 
\ee
Here $t = (T-T^*)/T^*$, (in the mean-field approximation used). In the pure limit $\kappa \to 0$ as $T^*$ is approached from above. With finite $<h^2>$, $\kappa$ does not $\to 0$ at any temperature, thus precluding true long range order. Note that $\kappa^2 = \kappa_0^2$ for $T \to 0$, where $t \to -1$. In (\ref{kappa}), the modification from the Ising model due to the fact that there are four states per unit-cell in the Ashkin-Teller model has been taken into account. In Fig. (\ref{fig:kappa}), $\kappa(T)$ is plotted to show its decrease as temperature decreases below $T^*$, for various values of $\kappa_0$. 

It should be noted that the linear in $T$ dependence of $S_e(q)$ in Eq. (\ref{Sq}) is valid only in the classical limit $T >> J \kappa^2$. For the size of the domains or $\kappa_0^{-1}$ larger than about 10 lattice constants, this is valid for $T$ larger than about $10^{-2} T^*$. For $T \to 0$, $T/(J\kappa^2)$ must be replaced by 1.

$S_e(q)$ and $S_c(q)$ for $q$ of order $\kappa$ will determine the  fermion self-energy in the next section. The relative contribution $S_c(\kappa)/S_e(\kappa)$ may be noted here. In the 'classical' case $(J\kappa^2/T) <<1$, the ratio, using the relation (\ref{kappa}) is,
\be
(J/T)(<<\Omega>_{av}>^2>_T\kappa_0^{-2})(<h^2>/J^2) = 4 \pi^3 (J/T) (T^*/J)^2 <<\Omega>_{av}>^2>_T,
\ee
For $<<\Omega>_{av}>^2>_T \approx 10^{-2}$, as estimated from polarized neutron scattering, this is of $O((J/T) (T^*/J)^2) >>1$, at $T<<J$ for $(T^*/J) \approx 2$. In the opposite or 'quantum' limit, $(J/T)\kappa^2)$ is replaced by 1 and an even larger ratio is obtained. Therefore in the self-energy calculations only 
$S_c(q)$ is used, especially as it also gives stronger variations in the self-energy. 

We may also roughly estimate $\kappa_0$. Such estimates can only be very approximate because one can guess only the order of magnitude of the parameters involved. For reasonable parameters, $c \approx 10^{-2}$, the local mean-square disorder energy $(\gamma <u>)^2$ of about $10J^2$,  $<\Omega_d>^2$ about $10 <<\Omega>_{av}>^2>_T$ at low temperatures, and other quantities as chosen in the previous paragraph, one finds (on recalling that $\kappa$ is in units of $2\pi/a$) that the correlation length of the domains, $\kappa^{-1}$ in real units is about 30 lattice constants, not too far from what is measured by STM. 

\section{Single-particle Spectral function}
To calculate the spectral function in the problem with domains, one needs, beside the distribution of the domains discussed in the last section, the single particle spectra of the pure problem in the loop ordered state and the scattering matrix due to the domains.

\subsubsection{One-particle Spectra in the Pure Loop-Current State}
A summary of the simple results \cite{simon} for the pure problem is given here. These simple mean-field calculations have been confirmed in detailed Monte-Carlo variational calculations \cite{weber}. The Hamiltonian $H_{{\bf \Omega}}$, the ground state, and the one-particle excited states and energies for each of the four order parameters ${\bf \Omega}$ in the pure limit are known. In mean-field approximation, 
 $H_{\bf \Omega}$ in the basis states $d_{\bf k}, p_{x{\bf k}}, p_{y{\bf k}}$ of the d-orbitals at the cu-sites and the oxygen orbitals on the x and y directions to it is given by \cite{simon}
 \be
  \label{H0}
H_{\bf \Omega} = \left(\begin{array}{ccc}0 & it_{pd}S_x & it_{pd}S_y \\-it_{pd}S_x & 0 &t_{pp}s_xs_y \\-it_{pd}S_y&t_{pp}s_xs_y& 0\end{array}\right)
 \ee

For simplicity the local energies at the three orbitals are taken to be degenerate, $t_{pd}$ is the hopping energy from the cu sites to the neighboring oxygen sites and $t_{pp}$ the hopping energy between neighboring oxygen sites, $s_{x} = sin(k_{x}a/2), s_{y} = sin(k_{y}a/2)$; $S_x = sin(k_{x}a/2 + \Omega_x), 
S_y = sin(k_{y}a/2 + \Omega_y)$. The order parameter ${\bf \Omega} = \Omega_x \hat{{\bf x}} + \Omega_y \hat{{\bf y}}$. $(\Omega_x, \Omega_y) = \Omega(\pm1, \pm1)$ specify the four different order parameters. 

For a given ${\bf k}$, the eigenvalues and eigen-vectors for $H_{{\bf \Omega}}$ depend on ${\bf \Omega}$. In the simplest approximation (in which $t_{pp} << t_{pd}$, the eigen-values of the conduction band, which alone will concern us in the rest of the paper, are
\be
\label{e0}
\epsilon({\bf k}, (\Omega_x, \Omega_y)) &\approx & t_{pd} \sqrt {S_x^2 +S_y^2},  \\
&\approx & t_{pd}\Big(\sqrt{\sin^2 (k_x a/2) + \sin^2 (k_y a/2)} + \frac{\Omega_x \sin(k_xa) + \Omega_y\sin(k_ya)}{2(\sin^2 (k_x a/2) + \sin^2 (k_y a/2))}\Big). 
\ee
for $|\Omega_x|, |\Omega_y| << 1$ (and excluding the unimportant region close to $k_x =0, k_y=0$). Changes due to the fact that $t_{pp}/t_{pd} \approx 1/2$ or that the three orbitals are likely to be not exactly degenerate do not produce any qualitative changes and considerably complicate the calculations. The four set of states for different ${\bf \Omega}$ are degenerate 
near $(k_xa, k_ya) =(0,\pi)$ and equivalents and maximally displaced in energy from each other near $(k_xa, k_ya)=(\pi/2,\pi/2)$ and equivalents. 
This will be important in giving the angular dependence of the pseudo-gap and the so-called "fermi-arc" phenomena. The eigen-vectors of the conduction band in the pure limit are,
\be
\label{ev}
|k, (\Omega_x, \Omega_y)\rangle = \frac{1}{\sqrt{2}}\big(i,~\frac{S_x}{S_{xy}},~ \frac{S_y}{S_{xy}}\big)^T.
\ee
where $S_{xy} = \sqrt{S_x^2 + S_y^2}$.

\subsubsection{Hamiltonian for Scattering between Domains}
In Fig.(\ref{domains}), space is divided into different regions; a particular $H_{\Omega_x, \Omega_y}$, reigns inside ${\cal{R}}_{\Omega_x,\Omega_y}$. The scalar product of the grounds state of the four different $H$ in the four different domains are of the order of the inverse of the area of a domains. The domain size is large compared to the area of a unit-cell so that such (Fock) states may to a good approximation be taken to be orthogonal and are therefore form a basis states for further calculations. For simplicity of notation, let us denote the states by $|{\bf k}, \alpha> $, where $\alpha = 1,..4$ labels $(\pm 1, \pm 1)$. In effect $\alpha$ serve as effective new quantum numbers between which there is forward scattering at the domain walls. In such a situation, the global ground-state and the wave-functions of excitations should be calculated such that they correspond to those of $H_{\alpha}$ inside each ${\cal{R}}_{\alpha}$ and transform continuously to those of $H_{\beta}$ on entering ${\cal{R}}_{\beta}$ due to the matrix elements of scattering between the domain $\alpha$ and $\beta$. 

The perturbation in traversing the domain between two domains is the difference  
\be
\label{pert}
H'({\cal R}_{(x,y)}, {\cal R}_{(x',y')}) = H_{\Omega_x', \Omega_y'} - H_{\Omega_x, \Omega_y},
\ee
with $H({\bf \Omega})$ given by (\ref{H0}). The basis state for the calculation are given by (\ref{ev}).

The matrix element of $H'({\cal R}_{(x,y)}, {\cal R}_{(x',y')})$  between the conduction band states given by (\ref{ev}),  are to first order in the $\Omega$'s independent of the momentum transfer ${\bf q}$:
\be
\label{matelem}
Lim~{\bf q} \to 0 M_{\alpha, \beta}({\bf k, k+q}) =  (t_{pd}/\sqrt{2}) \Big(\frac{(\Omega_x-\Omega_x') \sin(k_xa) + (\Omega_y-\Omega_y')\sin(k_ya)}{s_{xyk}}\Big) .
\ee
Here $ s_{xyk} \equiv \sqrt{s^2_{xk}+s^2_{yk}}$. Such matrix elements are to be taken at all the boundaries depicted in Fig. (\ref{domains}) through out the solid. We will have to consider $|M({\bf k,k})|^2$ in the calculation and integrate over the angle of ${\bf k}$ with respect to the domain boundary. This will yield a constant factor and an angle-dependent factor which integrated over the domain boundary and averaging over such domain boundaries yields a ${\bf k}$ independent contribution. So the only important thing about the matrix element is that it is finite for zero momentum transfer and its magnitude is $O(t_{pd})$ and that the scattering matrix is linear in ($\Omega_{\alpha} - \Omega_{\beta})$, with the order $\Omega_{\alpha}$ on one side of the boundary and $\Omega_{\beta}$ on the other.

\subsubsection{Procedure for the Calculation of the Self-Energy}
\begin{figure}
\includegraphics[width=0.8\textwidth]{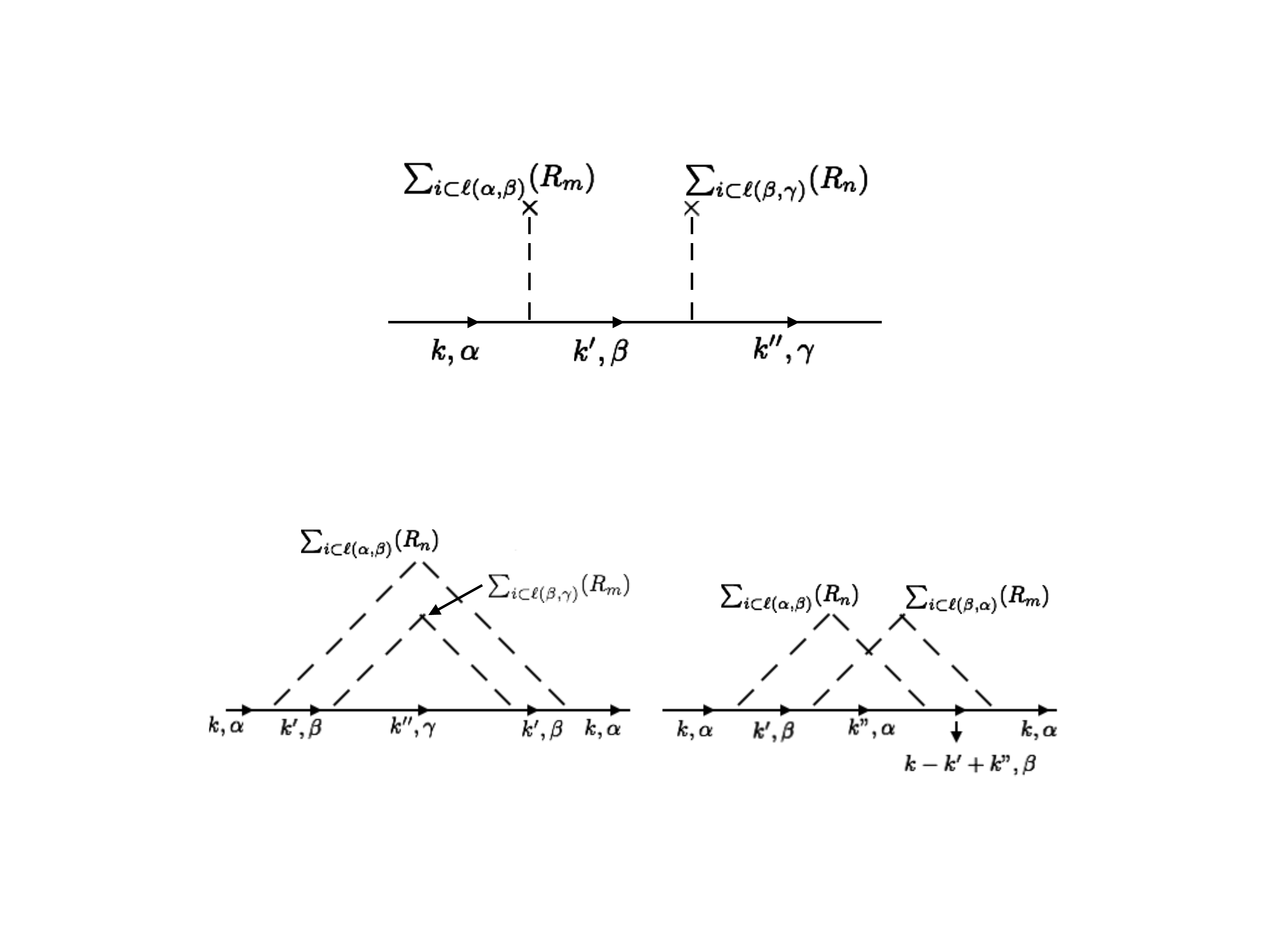}
\caption{Top: Bare impurity diagrams. $\sum_{i\subset \ell(\alpha, \beta)}(R_n)$ denotes the sum of all points $i$ which lie on the boundary between domains $\alpha$ and $\beta$ which lies at a location $R_n$, say the average position of the domain, in the lattice.  All $i$ in each such domain wall must be added coherently, while the location $R_n$ are randomly averaged. Bottom: Two "impurity"-averaged phase diagram. Each dashed line refers to the entire domain wall. For number of domains equal to N, the diagram on the right is $O(1/N)$ compared to that on the left even in the zero momentum transfer limit at each vertex because of the restrictions on the variety of the domains allowed on the right compared to those on the left.}
\label{imp}
\end{figure}
Consider the calculation of the self-energy when the 
domains of order, whose structure factor is given by Eq. (\ref{Sq}), are  arranged as in Fig. (\ref{domains}). This is an unusual problem of disordered fermions, not encountered to my knowledge before.
The calculation of the self-energy of fermions in this situation can be put into one to one correspondence with the diagrammatic technique developed for impurity scattering \cite{Edwards}\cite{AGD} with two important differences. \\

(1) One must consider the scattering from all points on  a given domain wall coherently but different domain boundaries scatter incoherently and are averaged assuming that they are located randomly with respect to each other.
Some "impurity"  diagrams for the self-energy are shown in Fig. (\ref{imp}). The scattering for a given domain wall common between $\alpha$ and $\beta$ has the matrix element
\be
\label{bare-scat}
\sum_{i \subset \ell(\alpha, \beta)}M_{\alpha, \beta}({\bf k, k}') e^{i({\bf k -k}').{\bf r}_{i, \alpha, \beta}},
\ee
where ${\bf r}_{i, \alpha, \beta}$ is the location of points on the domain wall $\ell(\alpha, \beta)$ between $\alpha$ and $\beta$. 
The averaging over the random location of different domain walls is done with (\ref{bare-scat}) as the vertex. On averaging,  domain walls $\ell(\alpha, \beta)$ over random locations in the sample, points on the domain wall scatter from $({\bf k}, \alpha) \to ({\bf k}', \beta)$ and back again to $({\bf k}', \beta) \to ({\bf k}, \alpha)$ as shown in the bottom two diagrams in Fig. (\ref{imp}). \\
(2) The other difference from the usual impurity scattering is due to the additional "quantum-numbers", between which the scattering takes place as noted in Fig. (\ref{imp}). Two diagrams of the same order, one "uncrossed" and the other "crossed" are shown at the bottom of Fig. (\ref{imp}). Usually crossed impurity self-energy diagrams are neglected for short-range scatterers because they are $O(1/(k_F \ell))$ smaller than the un-crossed diagrams; here $\ell$ is the mean-free path. This argument does not work for infinite-range or purely forward scatterers, $\propto  \delta({\bf k - k}')$ \cite{efros}, where crossed and uncrossed diagrams contribute equally. But, even for that case, as can easily be see, for example by comparing the two diagrams at the bottom of Fig. (\ref{imp}), that the crossed diagrams  are smaller than the un-crossed at each order by $O(1/N)$, where $N$ is the number of different domains. $N=4$ in the present case; we make use of this fact to neglect the crossed diagrams.

After neglecting the crossed diagrams and averaging over the random location of domain walls of density $c_d$,  the self-energy $\Sigma_{\alpha}({\bf k}, \omega)$ is given by
\be
\label{selfenergy}
\Sigma_{\alpha}(\fk, \om) \approx c_d<<\sum_{\bf k, k'}  \Big|\int_{\ell(\alpha,\beta)} d{\bf r}_{i} M_{\alpha, \beta}({\bf k, k+q})e^{i({\bf k-k}').{\bf r}_{i}}\Big|^2>_T>_{av} G_{\beta}({\bf k'}, \om),
\ee
where the averages are as defined earlier. $G_{\beta}({\bf k'}, \om)$ is the self-consistent Green's function for the $\beta$ domain. The contribution of different domains are added incoherently; hence the appearance of the density of domains $c_d$. The integral over $r_i$ is over the points in the domain wall $\ell(\alpha, \beta)$. We have already discussed that for to leading order $M_{\alpha, \beta}({\bf k, k}')$ may be taken at its value at momentum transfer ${\bf k-k'} =0$. Furthermore, one may make the approximation that on averaging over the randomly oriented boundaries, its ${\bf k}$ dependence may be neglected.  We are then left only with the spatial dependence of $\Omega_{i, \alpha} - \Omega_{j,\beta}$ which appears in the expression (\ref{matelem}) for $M_{\alpha, \beta}$. We have therefore to evaluate
\be
\label{domainwall}
<<\sum_{i,j \subset \ell(\alpha,\beta)}e^{i({\bf k-k}').({\bf r}_{i} -{\bf r}_j)} \Omega_{i, \alpha}\Omega_{j, \beta}>_T>_{av}, 
\ee
This then is simply the Static structure factor of the domain walls $S_w(|k-k'|)$, i.e. the Fourier transform of  $P_w(|{\bf r}_i-{\bf r}_j|_{arc})$, the probability distribution
 that two points on the domain wall between ${\cal R}_{\alpha}$ and ${\cal R}_{\beta}$ lie at the distance $|{\bf r}_i-{\bf r}_j|_{arc}$ measured along the domain wall.
 
 \subsubsection{Distribution of Domain Wall Lengths}
To find $P_w(|{\bf r}_i-{\bf r}_j|_{arc})$, we take the simple case that the domain wall boundary is the arc of a circle uniformly distributed over an angle from 0 to $2\pi$. Let $P(r)$ be the distribution of domain radii, which is actually the Fourier transform of $S(q)$. Then
\be
P_w(r_{arc}) = \int_0^{\infty}dr\frac{1}{2\pi}\int_0^{2\pi}d\theta \delta(r_{arc} - r\theta) P(r).
\ee
This gives that 
\be
 \frac{d}{dr_{arc}} P_w(r_{arc}) = P(r_{arc}/2\pi)
 \ee
The two contributions to the static structure factor $S_{e,c}(q)$ defined in Eq. (\ref{Sq}) give the radial distribution function at large distances,
$P_{(e,c)}(r) \propto \big((\kappa r)^{(-1/2)}, (\kappa r)^{(-3/2)}\big) e^{-\kappa r}$. Therefore it follows that the two contributions to the static structure factor for the domain walls, $S_{(e,w),(c,w)}(q)$ may be taken to be   $(\kappa)^{-1} S_{(e,c)}(q)$.

\subsection{Self-Energy}
The self-energy, Eq. (\ref{selfenergy}) a given domain $\alpha$ due to scattering at the boundary to states of the other domains $\beta$ is after these approximations,
\be
\label{se2}
\Sigma_{\alpha}(\fk, \om) \approx \frac{c_d}{(\pi/2)\kappa}\sum_{\bf q, \beta \ne \alpha} t_{pd}^2 ~S_{w}(\fq)~
G_{\beta}(\fk -\fq, \om).
 \ee
  This expression is reminiscent of the self-energy obtained long ago in Edwards' \cite{Edwards} derivation of the scattering rate in a liquid metal, where $S({\bf q})$ occurs as the structure factor of the liquid. Some differences from the usual impurity contribution to the self-energy may again be noted. The factor $(\pi/2)\kappa^{-1}$ is the average domain length, which must enter because of the sums in Eq. (\ref{domainwall}). Together with the factor $\kappa^{-1}$ coming from $S_w(q)$, discussed below, this sums over the scattering from a domain wall coherently. The factor $c_d$ reflects the sum over different domain walls, situated randomly in the sample incoherently. 

 The Green's function for the problem is obtained by averaging over the four $\alpha$'s. As a first step, one may use $G^0_{\alpha}(\fk, \om)$, the Green's function corresponding to $H^0_{\alpha}$. The self-consistent Green's function are found in one case by numerical iteration but no significant difference was found from the leading order answer. The results presented are the leading order results.

We must add to $\Sigma_{\alpha}(\fk, \om)$ the usual scattering due to short-range impurity scattering which gives an imaginary part due to scattering within each domain. This 
is, as usual nearly energy and momentum independent, featureless and may be included as a contribution $i/2\tau$. It is important to include it as it tends to limit the singular features due to (\ref{se2}). 

The imaginary part of the self-energy is calculated to be
\be
\label{ise}
Im \Sigma_{\alpha, \beta}(\fk, \om) &= &Im \Sigma_{e, \alpha, \beta}(\fk, \om)+ Im \Sigma_{c,\alpha, \beta}(\fk, \om). 
\ee
 $\Sigma_{e,c}$ are contribution due respectively of $S_{w,e}$ and $S_{w,c}$. The retarded part $Im \Sigma_{eR}$ is first found,
 \be
 \label{sigmac}
 Im \Sigma_{eR,\alpha, \beta}(\fk, \om) &=& Im \Sigma_{e0,\alpha, \beta}\sqrt\frac{(\kappa v_F )^2+ (\tau/2)^{-2}}{(\om-\xi^0(\fk,\beta))^2 + (\kappa v_F )^2+ (\tau/2)^{-2}} \\ \nonumber
 Im \Sigma_{e0,\alpha, \beta} &\equiv& - \frac{c_d}{2\pi\kappa^2}<<\Omega(T)>_T^2 >_d\Big(\frac{t^2_{pd}}{\kappa v_F}\Big)\frac{T}{J}.
 \ee
$\xi^0(\fk,\alpha) \equiv \epsilon^0({\bf k}, {\bf \Omega}_{\alpha}) - \mu$ are measured from the chemical potential (or equivalently from $\epsilon^0(\fk_{F}(\alpha))$. $Im\Sigma_{e0}$ is the value of $Im \Sigma_e(\fk, \om)$ at the fermi-surface and for $\omega =0$. 
The real part of the self-energy is obtained by Kramers-Kronig transformation and used in the calculations of the spectral function.

Using Eqs. (\ref{Sq}),  $\Sigma_c$ is found from $\Sigma_e$ by
\be
\label{sigmac2} 
\Sigma_{cR}(\fk, \om) = - \frac{J}{T} \frac{<h^2 >}{J^2}<<\Omega(T)>_T^2 >_d\frac{\partial}{\partial(\kappa^2)} \Sigma_{eR}(\fk, \om).
\ee
This is the easiest way to calculate $\Sigma_c$. As already discussed, for low temperatures $S_c$ is much larger than $S_e$. So only its contribution, $\Sigma_c$ is considered in the rest of the paper, especially as it produces sharper features in the spectral function than $\Sigma_e$.

\begin{figure}
\includegraphics[width=0.8\textwidth]{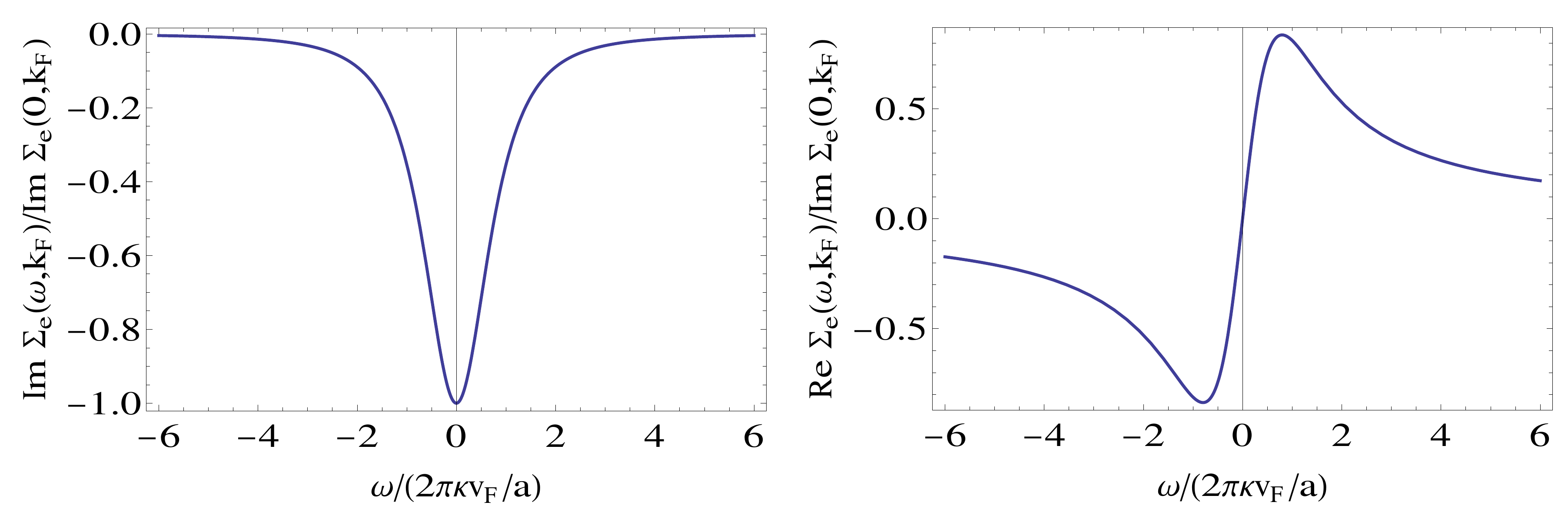}
\caption{The Imaginary and the Real part of the normalized Self-energy $\Sigma_e$, in the $(\pi,0)$-direction at the "Fermi-vector" as a function of energy
normalized with respect to the width parameter $(2\pi \kappa v_F/a)$. In this direction, the bare-energy $\xi^0(\fk,\alpha)$ are the same for all $\alpha$ so that domain averaging introduces no changes. $\tau^{-1} = (2\pi \kappa v_F/a)$ in this calculation.}
\label{Fig-ImReSE}
\end{figure}

The imaginary and the real part of the self-energy for the directions  $(\pi,0)$ and equivalent, where the bare energies for different ${\bf \Omega}_{\alpha}$ are the same are shown  in Fig. (\ref{Fig-ImReSE}). 

 \begin{figure}
\includegraphics[width=0.7\textwidth]{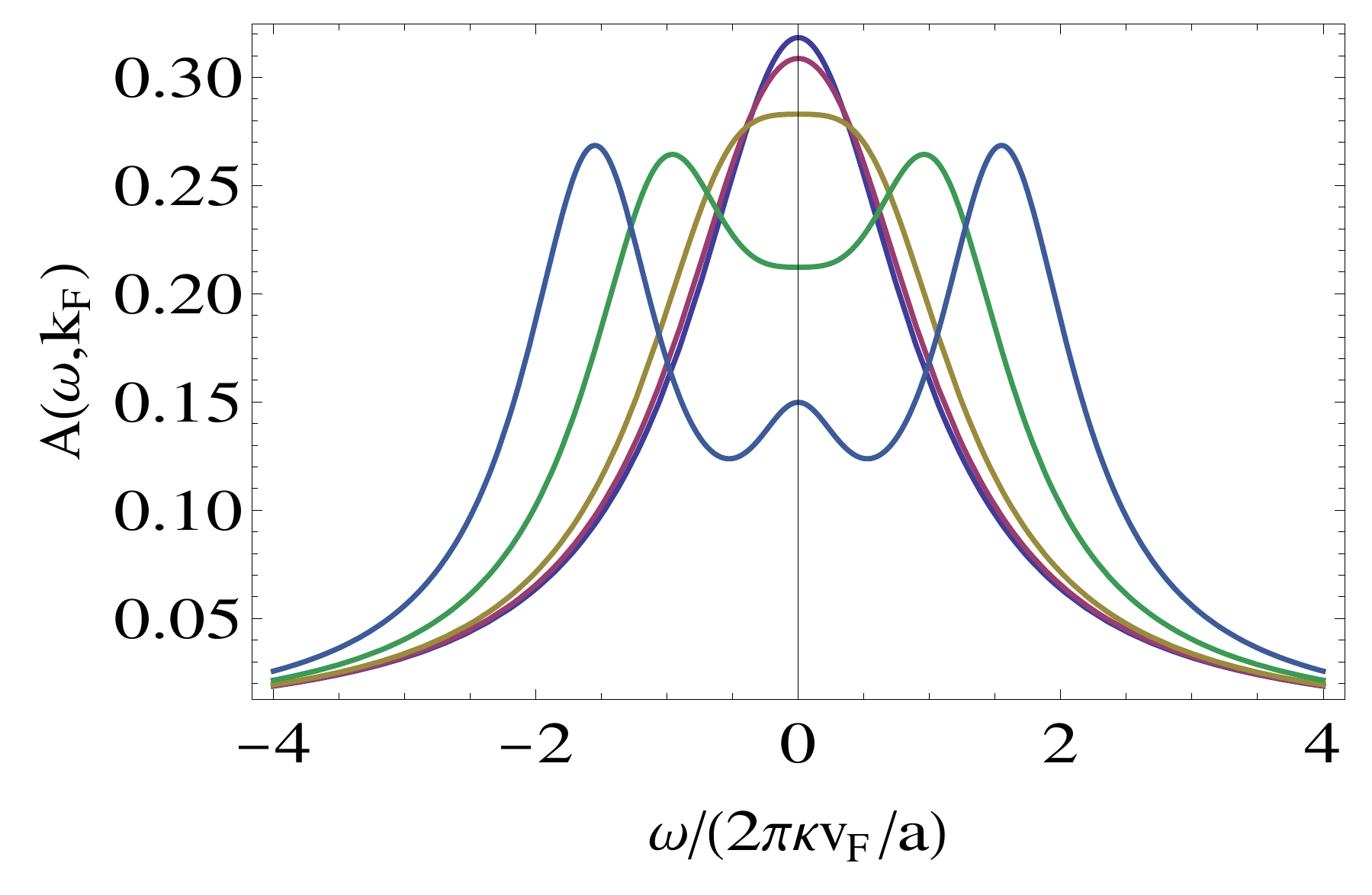}
\caption{The single-particle spectral function at the fermi-vector near the $(\pi,0)$ direction as a function of  energy in units of the width parameter $2\pi \kappa v_F(\theta)/a$. The successive curves  are for the coefficient Im $\Sigma_e(k_F,0)$ = 0, 0.25, 0.5, 1,  2 in units of 
$(2\pi \kappa v_F/a)$. This coefficient defined through Eq. (\ref{sigmac2})  depends on the order parameter and the Matrix elements, which increase on decreasing temperature. An impurity scattering contribution to the imaginary part of the self-energy $(2\tau)^{-1} = (2\pi \kappa v_F/a)$ has also been used in the calculations. The hump at $\omega=0$ for large order parameter disappears for larger impurity scattering.}
\label{Fig-A12e0}
\end{figure}
\begin{figure}
\includegraphics[width=0.7\textwidth]{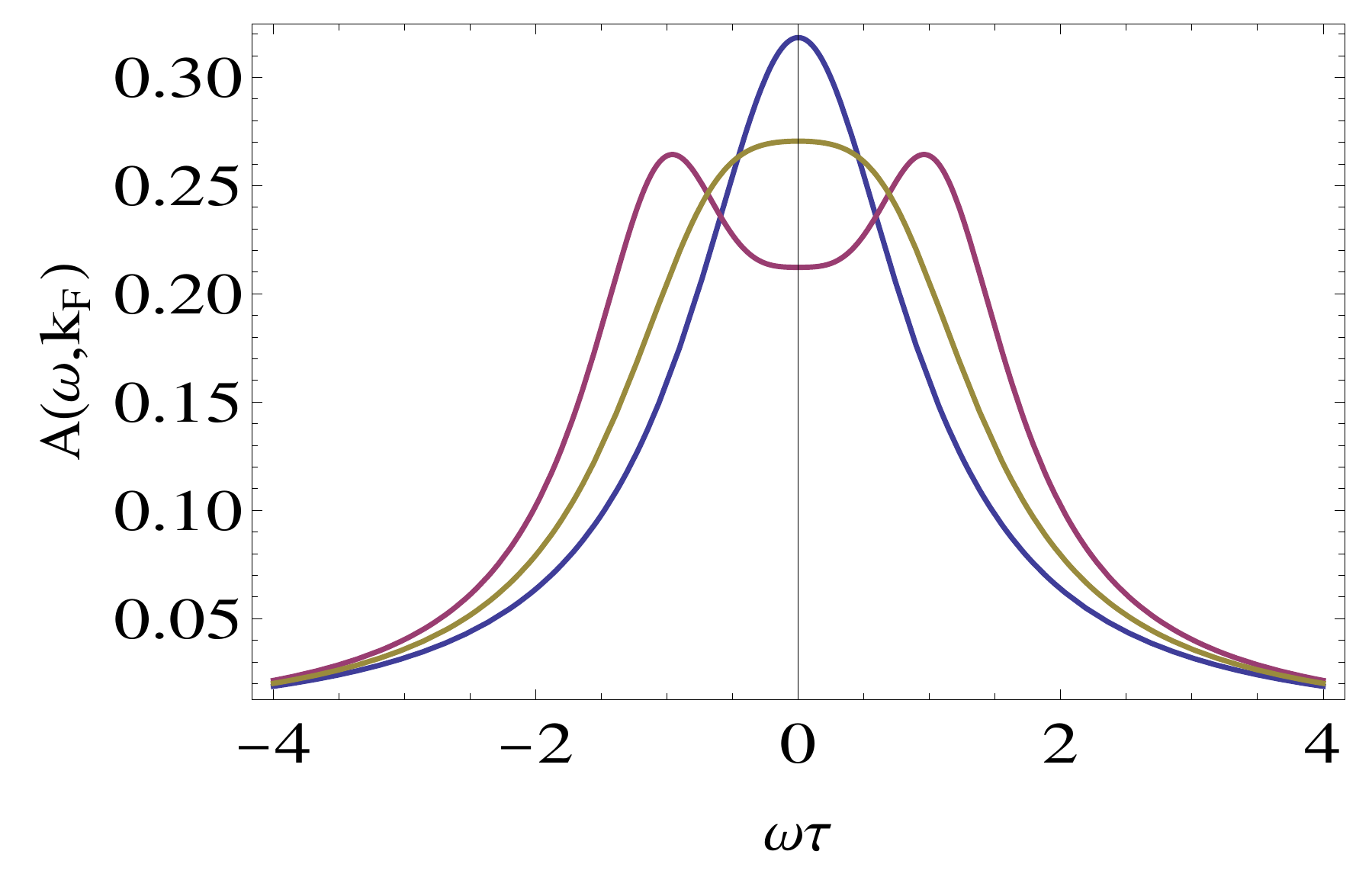}
\caption{The domain averaged single-particle spectral function at the fermi-vector as a function of  energy in units of a background scattering rate $\tau^{-1}$  for various values of  the normalized width parameter $2\pi \kappa v_F(\theta) \tau/a$ The successive curves are for $(2\pi \kappa v_F/a) =0.5, 1, 2$ in units of $\tau^{-1}$. $Im \Sigma(0,k_F) =1$ for all the curves.}
\label{Fig-Avskappa}
\end{figure}
\begin{figure}
\includegraphics[width=0.7\textwidth]{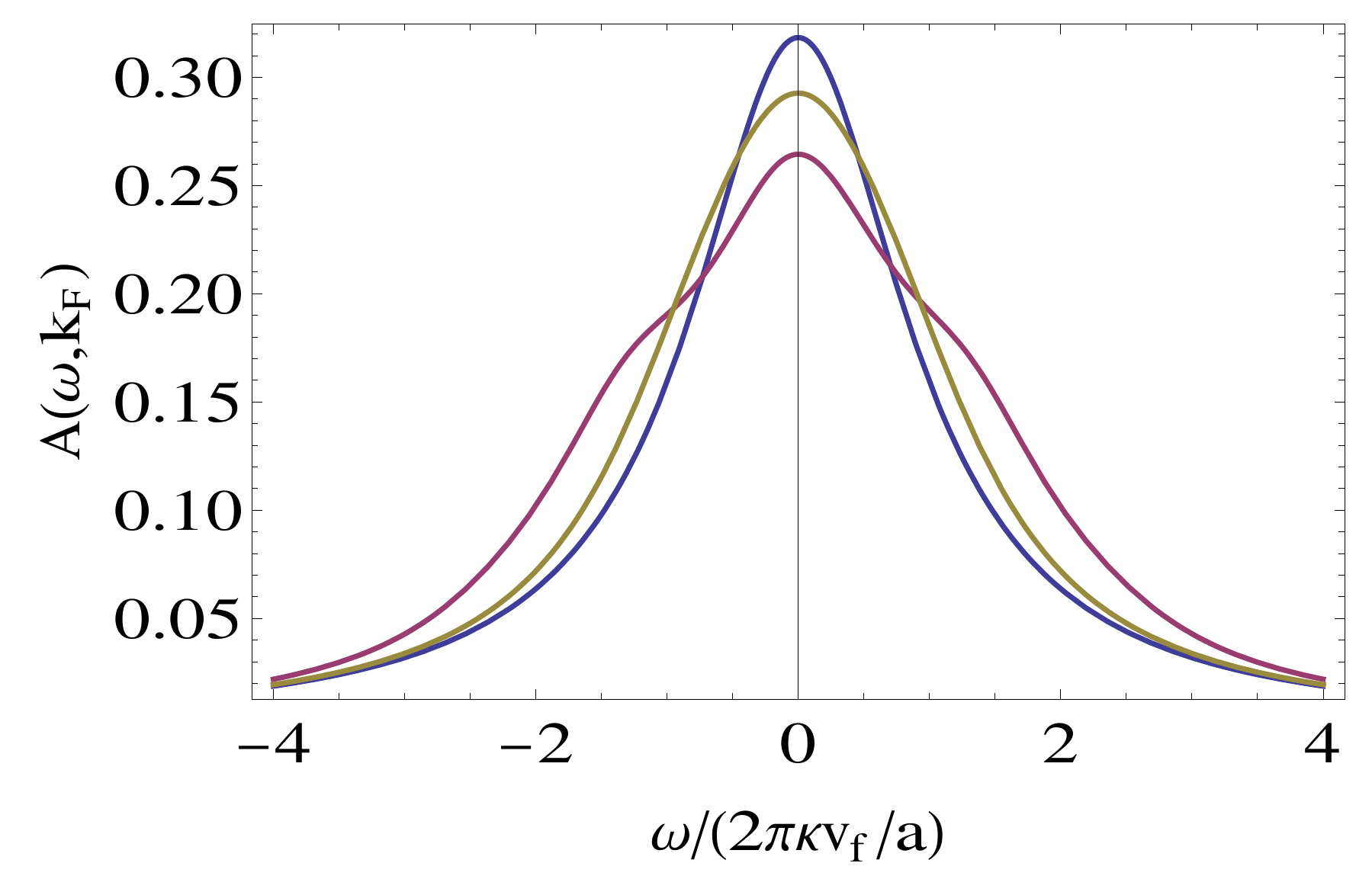}
\caption{The domain averaged single-particle spectral function at the fermi-vector in the $(\pi,\pi)$ direction as a function of  energy in units of a background scattering rate $\tau^{-1}$ in the $(\pi,\pi)$-directions for no loop current order and for increasing loop-current order, with Im $\Sigma_e(k_F,0)$ = 0, 1,  2 in units of 
$(2\pi \kappa v_F/a)$. The differences of energies for different ${\bf \Omega}_{\alpha}$ are correspondingly changed in each curve.}
\label{Fig:pi-piSpecFn}
\end{figure}
\begin{figure}
\includegraphics[width=0.7\textwidth]{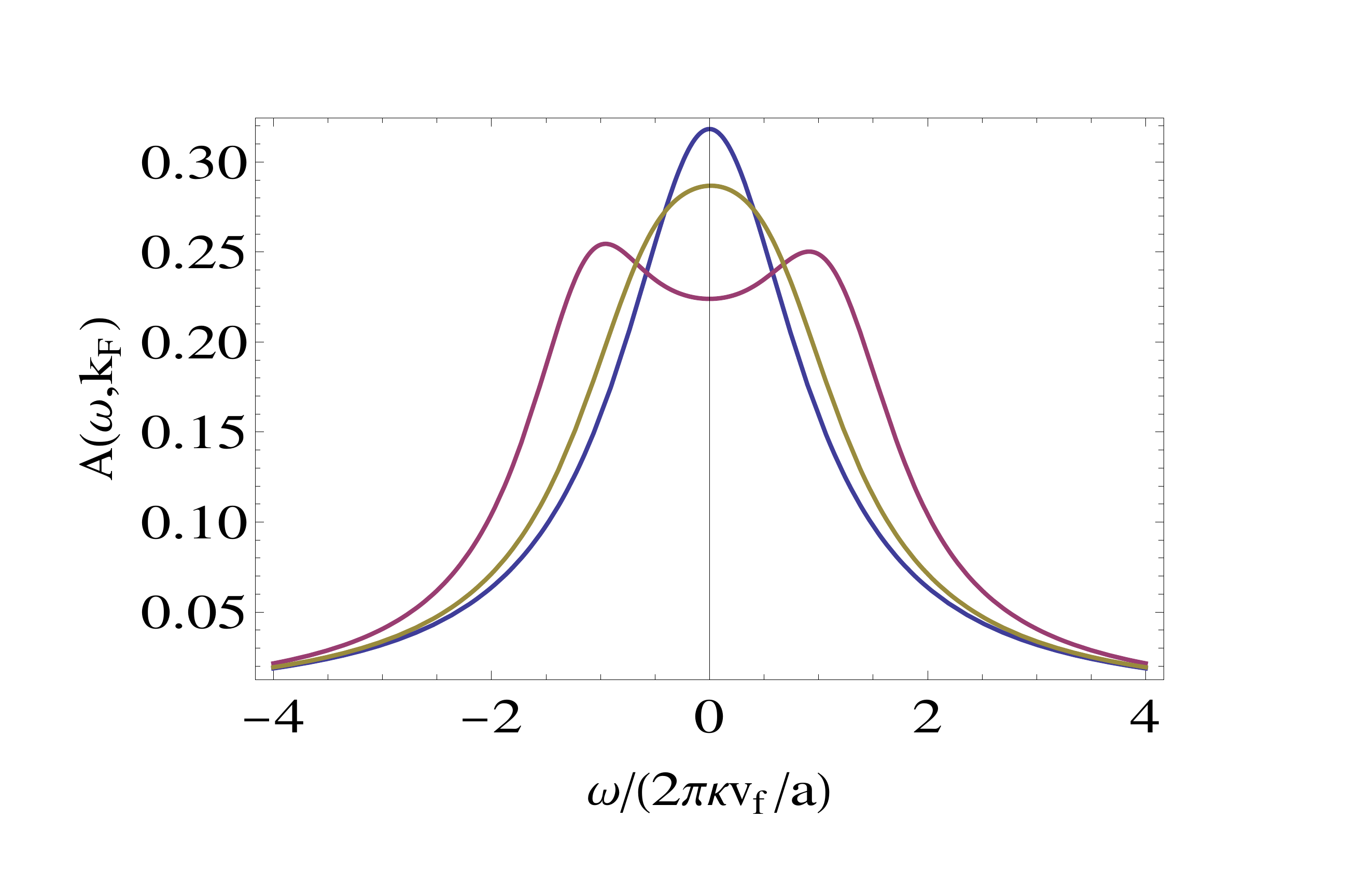}
\caption{The domain averaged single-particle spectral function at the fermi-vector in a direction $\pi/8$ with respect to a crystalline axis as a function of  energy in units of a background scattering rate $\tau^{-1}$  for no loop current order and for increasing loop-current order, Im $\Sigma_e(k_F,0)$ = 0, 1,  2 in units of 
$(2\pi \kappa v_F/a)$. The differences of energies for different ${\bf \Omega}_{\alpha}$ are correspondingly changed in each curve.
The magenta curve is at the same value of $|\Omega|$ as the blue curve in the figure for the $(\pi,\pi)$-direction and the green curve for the $(\pi,0)$-direction.}
\label{Fig-Api/8}
\end{figure}

The {\it domain averaged} spectral function is calculated using the approximations described above,
\be
\label{spfn}
A(\fk, \om) = -\frac{1}{\pi} Im G_R (\fk, \om) \approx -\frac{1}{4\pi} \sum_{\alpha}Im \frac{1}{\om -\xi^0(\fk,\alpha) - \Sigma_{\alpha, R}(\fk, \om)}.
\ee
Here $\Sigma_{\alpha, R}(\fk, \om) \equiv \sum_{\beta \ne \alpha}\Sigma_{\alpha,\beta, R}(\fk, \om)$. The results for $\xi({\bf k}, \alpha)=0$ in (or close to) the $(\pi,0)$ directions are shown in Figs. (\ref{Fig-A12e0}) for various values of the pre-factor of the self-energy $Im\Sigma({\bf k}_F,0)$.  In Fig. (\ref{Fig-Avskappa}) the sensitivity of the spectral function to the width parameter $(2\pi \kappa v_F/a)$ is exhibited.
(In all the figures the physical dimensions of $\kappa$ are restored.)
The results for $\xi({\bf k}, \alpha)=0$ in the $(\pi,\pi)$-directions are shown in Figs. (\ref{Fig:pi-piSpecFn}). As explained above the contributions of the spectral function for different ${\bf \Omega}_{\alpha}$ have been summed after the self-energy for each of them to scatter to the others in the adjoining spatial regions is summed. One of the curves is drawn with the matrix element for scattering set to zero representing the spectral function without loop order and the other two are for increasing magnitude of loop-order.  
 To see the passage from the $(\pi,\pi)$ direction to $(\pi,0)$ direction, the spectral function is plotted for the same three coupling constants and values of $\Omega_{\alpha}$ at the angle bisecting $(\pi,\pi)$ direction to $(\pi,0)$ directions, in Fig.( \ref{Fig-Api/8})

\subsection{Features of the Calculated Results and Comparison with Experiments}
Let us now compare features of the derived pseudo-gap  to the 
ARPES experiments which in principle give the most detailed information about the pseudo-gap. In doing so, one should note that there are no detailed analysis of the line-shapes observed in the experiments or extracted self-energies from which quantitative features of the data may be compared in detail with the calculations. Indeed, most of the data is only analyzed to get the peak position as a function of angle on the fermi-surface and the temperature. I suggest that the self-energy below $T^*(x)$ be deduced from experiments and compared with the expressions and calculations give here to test their validity. However several general qualitative and some quantitative features may already be compared.
  
 (i) {\it Magnitude of the gap}: The position of the principal peaks in the spectral weight for ${\bf k} ={\bf k}_F$ and in the anti-nodal direction with respect to the chemical potential, i.e. the pseudo-gap G may be estimated from the self-energy, Eqs. (\ref{sigmac}, \ref{sigmac2}). It is easy to see that there can be no discernible gap with magnitude much smaller than $O(\kappa v_F, \tau^{-1})$. Let us consider the opposite possibility that  $G >> \kappa v_F, \tau^{-1}$ is realized. In that limit, we may replace the factor $(\kappa^2 + q^2)^{-2}$ in $S_c(q)$ by $\approx \kappa^{-2} \delta(q^2)$. Then, it is easy to see that the real part of the self-energy for the single-particle Green's function at ${\bf k}$ at the Fermi-vector ${\bf k}_F$ in the direction in which two domains have identical energy is  $\propto (\omega -\zeta({\bf k}_F))^{-1}$, which allows an easy estimation of $G$ by looking at the poles of 
 \be
 \frac{1}{\omega -\zeta({\bf k}_F) - \frac{E_0^2}{(\omega -\zeta({\bf k}_F))}}.
 \ee
 $E_0$ is given by Eqs. (\ref{sigmac}, \ref{sigmac2}), so that there is  gap at the chemical potential in that direction 
  \be
   \label{G}
G \approx  2E_0 \approx  2\Big[\frac{c_d}{\kappa^2} ~ \frac{c}{\kappa^2}\frac{<h^2>}{J^2} \big(<<\Omega(T)>_T^2 >_{av} \big)^2(t^2_{pd})\Big]^{1/2}.
  \ee
$c_d/\kappa^2 \approx 1$, since the domains cover the sample and the inverse of the typical area of a domain is $\approx \kappa^{-2}$. Using the same values of the parameters as at the end of Sec. II B, which gave size of the domains $\kappa^{-1}$ in real units to be $\approx 30$ lattice constants gives $G \approx 0.1 eV$ for $t_{pd} \approx 1 eV$. This value of $\kappa$ gives $\kappa v_F \approx 0.06 eV$. So $|G|$ is of the same order as  $2\pi\kappa v_F/a$ and the limiting procedure used here does not work very well. Keeping the imaginary parts due to $\kappa v_F, \tau^{-1}$ is essential. This is done in the numerical results presented above where the the self-energy and the spectral function are presented with energy on the scale of $2\pi\kappa v_F/a$, (using real units for $\kappa$).  In the limit used to get Eq.(\ref{G}), the impurity dependence is $\propto c^{1/2}$. More importantly, we find from the numerical calculations that the conclusion from the approximate argument made here that the magnitude of the gap
is proportional to  $\propto <<\Omega(T)>_T^2 >_{av}$ holds approximately in the numerical calculations.
\begin{figure}
\includegraphics[width=0.7\textwidth]{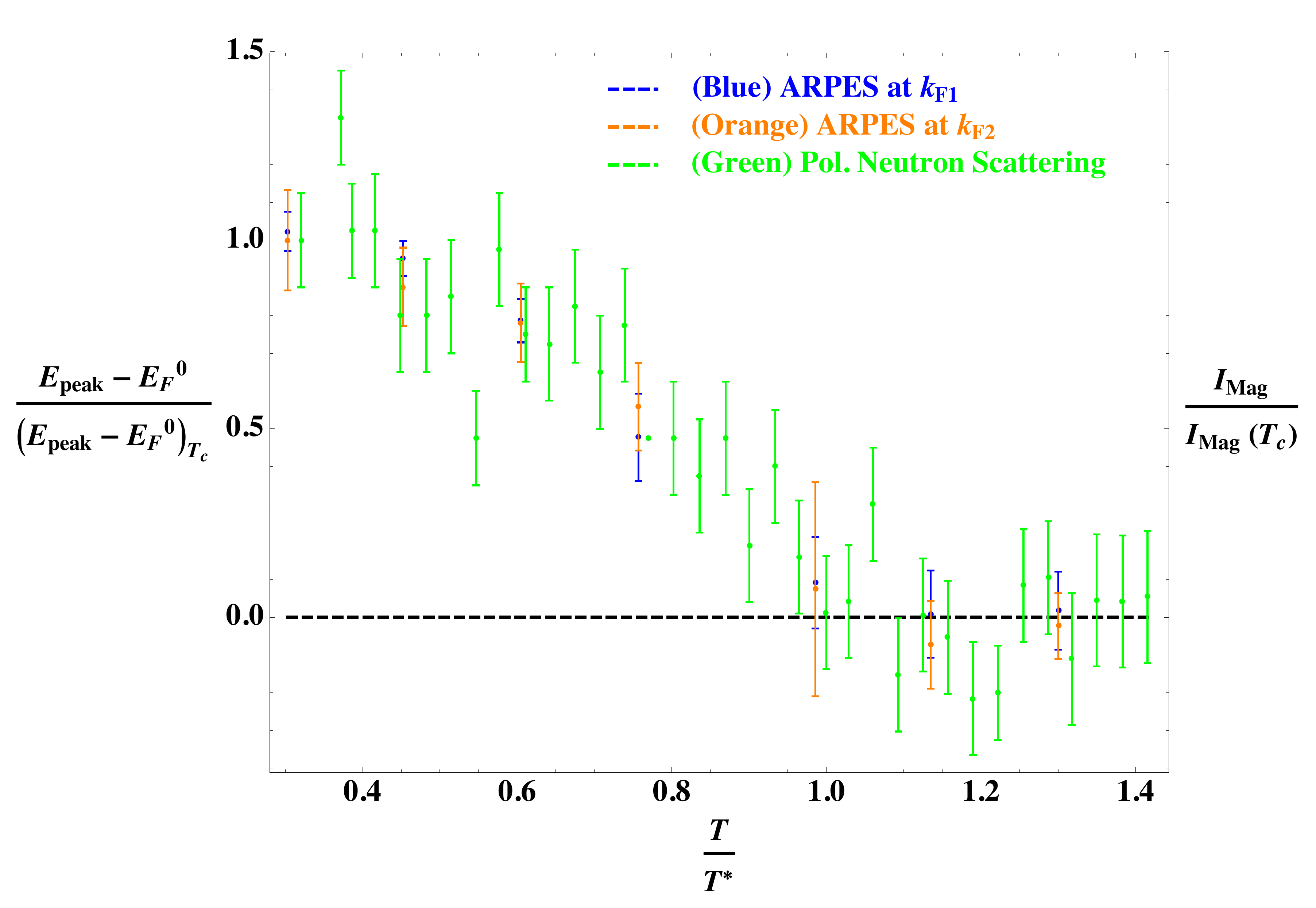}
\caption{The data for the movement of the position of the peak in the spectral function for Bi2201 as a function of temperature, Fig.(S1-F)  from Ref(\onlinecite{He-arpes-kerr}) from above $T^*$ to $T_c$ is plotted in a normalized to its value at $T_c$ as a function of $T/T^*$. The measured magnetic Bragg peak intensity at $(101)$ measured by polarized neutrons in Ref (\onlinecite{greven-prb}) in $HgBa_2CuO_{4+x}$ from above $T^*$ to near $T_c$ is similarly normalized and plotted as a function of $T/T^*$. The ratio $T_c/T^*$ for these two compounds is the same to within about 10\%. $T^*$ in both sets of data also have an uncertainty of about 10\%}
\label{Fig-Arpes-neutrons}
\end{figure}
 
(ii) {\it Temperature and Angular Dependence of the Pseudo-gap - ``The Fermi-Arc"}:   In Fig. (\ref{Fig-A12e0}), the spectral function is plotted for several values  of $Im \Sigma(0, \hat{{\bf k}}_F)$ in the $(\pi,\pi)$ direction. Since the latter is proportional to $<<\Omega(T)>_T^2 >$, the curves can be interpreted as the spectral function as a function of temperature. 
 $<<\Omega(T)>^2>$ can be obtained from the polarized neutron scattering intensity, for which the results are available for $YBa_2Cu_3O_{6+x}$ \cite{bourges} and for $HgBa_2CuO_{4+x}$ \cite{greven-prb} while the information about the pseudo-gap is obtained in $Bi(2201)$ and $Bi(2212)$. Within the (rather large) error bars in the experimental results, the normalized Intensity of neutron scattering vs. the normalized temperature $T/T^*(x)$ of all the four measured samples for different $x$ in the latter and all the six measured samples in the former fall on top of each other. In Fig. (\ref{Fig-Arpes-neutrons}), the experimental data, taken from Ref. (\onlinecite{He-arpes-kerr}) for the temperature dependence of the deviation of the peak in the spectral function from the chemical potential at ${\bf k} ={\bf k_F}$ in two different directions is plotted in normalized units as a function of $T/T^*$, where $T^*$ is determined only to an accuracy of about $\pm 20 K$. On the same plot, the normalized square of the order parameter, measured by neutron scattering for $HgBa_2CuO_{4+x}$ is plotted.
In the limit that (\ref{G}) has been evaluated the dependence of the gap should indeed be as the (order parameter)$^2$, (assuming that the defective cells have a similar temperature of the order parameter as the typical cell). Within the uncertainty of the results shown in Fig. (\ref{Fig-Arpes-neutrons}), the dependence calculated is consistent with the observations. 

The effect of the correlations is  felt at an angle $\hat{{\bf k}}_F$ only when $G(T, \hat{{\bf k}}_F) $ is comparable or larger than the characteristic width $(2\pi v_F/a)\kappa(T)$. This is clearly in evidence in Figs. (\ref{Fig-A12e0}) and (\ref{Fig-Avskappa}). This automatically produces the phenomena termed "Fermi-arcs", which is simply the result that the spectral function as a function of angle $\hat{{\bf k}}_F$ from the nodal direction has a peak at the chemical potential for any ${\bf k}$ only over some angles near the nodal direction which decreases as temperature decreases. On comparing Figs. (\ref{Fig-A12e0}), (\ref{Fig:pi-piSpecFn}) and (\ref{Fig-Api/8}) for the $(\pi,0)$, $(\pi,\pi)$ and the direction at an angle bisecting them for the same values of $|\Omega|$, one finds that over this range, the gap is larger for the first than the third and non-existent for the second.  As temperature decreases below $T^*$ and $\Omega$ increases, the pseudo-gap appears first in the $(\pi,0)$ directions and progressively moves towards the $(\pi,\pi)$ directions. This naturally simulates the appearance of a Fermi-arc.

 There is always finite weight in the spectral weight at the chemical potential ($\omega= 0$) at $\xi({\bf k})=0$, but which is reduced to $Im\Sigma^{-1}(0,k_F)$ with a width which is proportional to inverse of its height. For large value of the pseudo-gap, a weak peak is displayed at $\omega =0$ in Fig. (\ref{Fig-A12e0}). Whether the peak occurs or not depends on details such as the scattering rate due to the usual impurity contribution and inelastic and temperature dependent relaxation rates. 
 
 While as noted, the average pseudo-gap depends weakly on the average impurity concentrations, locally the structure factor depends linearly on the impurity concentration and its relevant potential. Also, there is the effect of impurities which couples to the magnitude of loop-order. So, one should expect local gaps to be larger in the neighbor-hood of impurities. This has been observed in STM measurements \cite{stm-rev}, which shows larger gaps in the vicinity of identifiably impurities with the correlation length of the "gap-map" similar to that of the distribution of impurities.
 
\begin{figure}
\includegraphics[width=0.8\textwidth]{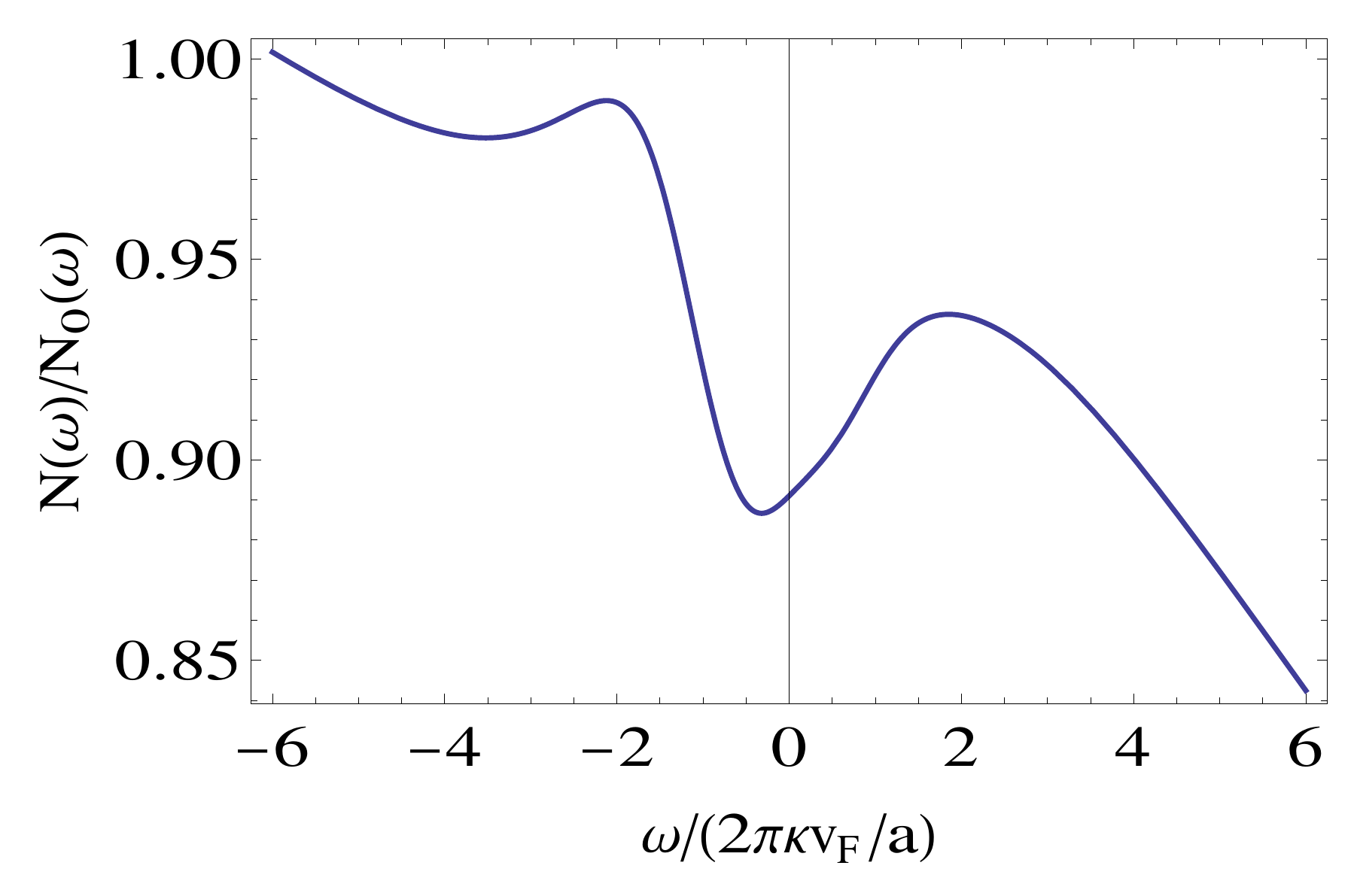}
\caption{ The density of states $N(\omega)$, normalized to its constant assumed value without the pseudo gap, is shown as a function of  energy $\omega$ in units of the width parameter $2\pi \kappa v_F(\theta)/a$. Impurity scattering rate is taken as 1 and  $Im \Sigma^{(1)}(\hat{{\bf k}}_F,0)$ = 1 in units of the width parameter.}  
\label{Fig-DOS}
\end{figure}

(iii)  {\it Density of States}: Scanning Tunneling Spectroscopy (STM) measures the angle-integrated density of states as a function of energy $N(\omega)$ multiplied by tunneling matrix elements which in general are functions of angle for which no precise information is possible. With this reservation STM provides very accurate density of states both below and above the chemical potential as well as its variation from one region to another on unit-cell size or better length scale.
A feature found in these experiments both above and below $T^*(x)$ is a decrease in the background density of states from below to above the chemical potential. Since, it is quantitatively similar above and below $T^*(x)$, it cannot have to do with the physics of the pseudo-gap. Over the range -100 meV to +100 meV, the decrease in density of states is about 25\% of the density of states at the chemical potential.  The simplest reason for the variation of the background density of states must be sought in the band-structure before any more sophisticated explanation are invoked. For tight-binding one-electron models which fit the experimentally determined fermi-surface for a Bi2212 sample with 16\% doping, which is in the pseudo-gapped region for which STM and ARPES results are available, the density of states has been calculated by Norman \cite{Norman-pc}  and shows variation similar to the background variation in the experiments.
I plot in Fig. (\ref{Fig-DOS}), the calculated density of states with such a background variation showing the pseudo-gap, which is about the scale of the depression of density of states at $\omega =0$ relative to the background which is observed with the pre-factor in self-energy indicated. 
 
(iv) {\it Upper Energy Scale of Change of Density of States}:  \\
 An interesting general aspect of the self-energy from the forward scattering process revealed in Eqs. (\ref{ise}) is that the imaginary part is a homogeneous function of $(\omega -\xi({\bf k}))$, so that as a function of this quantity it is important throughout the band. The real part  well below the chemical potential, $\omega <<0$ (or well above) decays very slowly as $\log|\xi({\bf k})/\omega_c|$ for positive $\omega = \xi({\bf k})$.  
 
The (almost) homogeneous in $(\omega -\xi({\bf k}))$ leads to anomalies in the spectral function tied to $\xi({\bf k})$ irrespective of the value of  $\xi({\bf k})$ within the band. This leads all states above the chemical potential to move up  and all states below the chemical potential to move down in energy, leaving a gap at the chemical potential. 
There should be anomalies at band-edges but these may be wiped out by additional sources of line-width, either intrinsic or inherent in the experimental tool employed. Forward scattering singularities imply that in measuring physical quantities, for instance the entropy or the single-particle density of states or the optical spectral weight, one must integrate over temperatures or frequencies over the scale of the band-width to satisfy conservation rules for the particle number and its consequences for such properties. This is to be contrasted with $\omega$ dependent but $\xi({\bf k})$ independent features of the marginal fermi-liquid form \cite{mfl} and for heavy-fermions \cite{cmv-hf}, \cite{george-dmft}, where the anomalous features are tied to the chemical potential. 

This form of the change in density of states explains the remarkable feature noted in the specific heat results  \cite{loram} and the c-axis optical sum-rule \cite{timusk-rev}, \cite{basov}, \cite{bontemps}. In the specific heat measurements which extend from low temperatures to just above $T^*(x)$ the entropy is not conserved compared to those in dopings without the pseudo gap.  The same anomaly is found more dramatically in the deduced c-axis optical spectral weight where f-sum-rule is not satisfied when integrating up to energies up to an order of magnitude larger than the pseudo-gap energy scale. (It is harder to deduce this from a-b plane conductivity, firstly because unlike c-axis conductivity, it is dominated by the region of states near the node due to their higher velocity \cite{bontemps} and also because it cannot be measured as accurately as the c-axis conductivity.) These results are consistent with STM measurements which directly show a diminution in density of states near the chemical potential without a bump in the density of states at larger energy needed to conserve the particle number when integrating it over a few times the gap energy. 
The same behavior has been noted recently in
ARPES measurements \cite{hashimoto2014}, where pseudo-gap is shown to lead to a change in density of states at energies over the entire range measured, of about 200 meV below the chemical potential. 

This form of density of states change is quite unlike what is seen by STM and ARPES on going below $T_c$, where the d-wave BCS singularity at $\pm\Delta_0$ and the satisfaction of the sum-rule within $\omega \approx 4 \Delta$, are found. For CDW or AFM transitions in metal, satisfaction of the sum-rules over energies a few times the gap is also the general rule.

The almost homogeneous dependence of the spectral function on $(\omega-\xi_{\bf k})$ may be tested experimentally by ARPES. However, it should be remembered that there should in addition be a contribution to the self-energy at temperatures and frequencies  above $T^*(x)$ of the marginal fermi-liquid form, which is almost purely $\omega$-dependent, as is indeed observed \cite{lijun} \cite{rameau}. 

(v) {\it Low energy transport measurements}: There is a finite density of states at the chemical potential and for energies below the pseudo-gap scale $G$.  The singularities responsible for the marginal fermi-liquid anomalies in the quantum-critical region are also removed in the low-energy spectra of the loop-ordered state. So if one confines experiments to low energy and temperature compared to $G$, deceptively simple normal fermi-liquid behavior is to be expected. This has been well documented in various experiments \cite{Barisic} \cite{vandermarel-1}. Theoretical calculations for transport properties in the pseudo-gap regime using the remnant "Fermi-Arc" in agreement with experiments have been presented \cite{gorkov-fermiarc}

(vi) {\it Particle-hole asymmetry}: We note that the spectra is not particle-hole symmetric {\it in detail} even when we assume a constant normal density of states over the pseudo-region of energies. Evidence for this is presented in the calculated density of states, Fig. (\ref{Fig-DOS}). Similar asymmetry is calculated when only a range of $\ek$ on either side of the normal ${\bf k}_F$ is integrated over or if the spectal function is compared for $\pm |{\bf k}-{\bf k}_F|$. The diminution in spectral weight above the chemical potential is smaller than that below in such measurements. This asymmetry was first observed in Ref. (\onlinecite{johnson-assymm}). In the calculations here, the asymmetry comes about from the analytic properties of the real part of the self-energy for forward scattering singularity, which insist that it be zero in the limit $\omega \to -\infty$ from negative values, in the limit $\omega \to +\infty$ from positive values, be zero at $\omega = 0$ and have sharply varying behavior near $\omega =\xi({\bf k})$.

(vii)  {\it Frequency Dependent Susceptibility: Effects on Local Probes}:\\
 It is known from experiments in random-field Antiferromagnets \cite{birgeneau} that time-dependent fluctuations in such problems are very complicated \cite{v}, \cite{nattermann},  \cite{dedominicis}, similar to that in spin-glasses. Experiments with different time-scales have given different results. Specifically in relation to the dichotomy in the results from "elastic" neutron scattering and NMR probes in cuprates mentioned in the Sec. I, there are instances such as in UPt$_3$, where neutron scattering found long-range order \cite{upt3-neutron} but NMR did not find an onset of static fields\cite{upt3-nmr} below the AFM transition. There are two distinct aspects of the time-dependence. One is the complicated problem of the dynamics of the domain walls \cite{dedominicis} which also leads to hysteretic phenomena. These are in general very slow phenomena and not of our concern for the present problem. The other is that of the characteristic scale of  fluctuations  of lengths smaller and up to order of their finite size. Such fluctuations at low temperature are dominated by quantum fluctuations. 
    
Then derivation of the static structure factor \cite{Toulouse, HV, cowley, dedominicis} , summarized in Sec. II, assumes that the dynamic fluctuations are of the Orenstein-Zernike form 
\be
\label{chiqw}
\chi_c(q,\omega) = \frac{<<\Omega(T)>^2 >}{i \omega+ J(q^2 + \kappa^2)}. 
\ee  
At long wavelengths, $q \lesssim \kappa$, the characteristic frequency of oscillations never truly go to 0 at any temperature because $\kappa(T)$ never approaches 0. For larger $q$, the dynamics well below $T^*$ must revert to that of the pure system, i.e. finite energy oscillations between different possible configurations of the order parameter. These have been observed \cite{Li-collmodes} and extensively investigated theoretically \cite{he-cmv-collmodes}.
Eq. (\ref{chiqw}) says that there must exist low frequency excitations with a characteristic frequency width of order
  \be
  \label{dom}
  \Delta \omega \approx J\kappa^2(T).
  \ee
The lower limit to the correlation length $\xi= (2 \pi \kappa)^{-1}a$ through neutron scattering measurements in the crystal studied by neutron scattering is about $25 a$.  With the characteristic nearest neighbor or zone boundary energy scale $J/(a/(2\pi))^2 \approx 500$ Kelvin, i.e. $10^{13} sec^{-1}$ this gives $ \Delta \omega $ less than about $10^{10} sec^{-1}$. This is  much smaller than the characteristic resolution of the neutron scattering experiments which for "quasi-elastic peaks" integrate over energies of $O(1 meV) \approx 10^{11} sec^{-1}$, while it is much above the typical NMR shifts expected which are $ O(10^5) sec^{-1}$. 
Static fields in different directions and with varying magnitudes are then motionally averaged \cite{NMRbook} due to the fluctuations at frequencies much larger than the NMR or $\mu$SR shifts or expected in the pure model.  
Motional averaging occurs if the fluctuation frequency between alternate configurations, $ \Delta \omega$ is much larger than the energy splitting $\delta \nu$ between alternate configurations or the frequencies over which they are distributed. Only the un-split line can then be seen. The contribution to the line-width of this process is $ \approx (\delta \nu)^2 ( \Delta \omega)^{-1}$. For $\Delta \omega \approx 10^{10}$secs$^{-1}$ and 
$\delta \nu \approx 10^{5} $secs$^{-1}$, this is $\approx 1 $sec$^{-1}$.  If there are independent sources of line-width much larger than this, say direct nuclear relaxation through conduction electrons or dipolar interactions, then this line width is unobservable in $T_1^{-1}$ measurements. For NMR relaxation in $YBa_2Cu_3O_{6.63}$, the nuclear relaxation rate \cite{walstedt} of $^{17}O$ at the chain sites at 200 K is about 40 sec$^{-1}$. The NQR experiment \cite{nqr} on $^{137}$Ba shows a relaxation rate \cite{nqr} of about $3$ X $10^{3} secs^{-1}$. (There are some special aspects of the NQR experiments \cite{nqr}, which are discussed in a separate paper \cite{hou}.) In experiments with muons, the relevant time scale is muon lifetime itself, which is only 2 $\mu$secs. So motional narrowing leaves in every case an unshifted line with no traces left of the shift or the motionally averaged relaxation rate. However, if there are nuclei with relaxation rate much slower than a few secs$^{-1}$, one should observe a change of relaxation rate due to the motional narrowing discussed here.

One can also ask, what would the correlation length have to be in order to see line-splittings and shifts similar to static long-range order, i.e. when is the characteristic scale of fluctuations smaller than $10^5 sec^{-1}$. The answer from the above estimates is $\xi \gtrsim 10^4~a$, i.e. O(4 micron).

These consideration suggest a way to resolve the apparent conflict between the observed order in neutron scattering and in low frequency local experiments. The idea can be tested in further experiments which span the range of frequencies of the two kinds of experiments. Some evidence of new relaxation rates setting below the pseudo-gap temperature has been obtained in pump-probe spectroscopy \cite{orenstein}. There also appear to be some abnormal low frequency relaxational processes \cite{Gedik}
and low frequency microwave absorption up to the range of about 10 GHz \cite{bonn} in under-doped cuprates, which appear to disappear for dopings without a pseudo-gap. These need further investigation in light of the results presented here.

\section{Further Tests of the Theory and Concluding Remarks} 
The most important suggestion for the experiments is a detailed investigation of the ($\omega, {\bf k}$)-dependence of the self-energy to test
whether it is consistent in detail with the form derived in Sec. (IV) here. In particular the predicted $(\omega-\xi({\bf k}))$ dependence is  unique to forward scattering. These can be deduced by quantitative measurements using the ARPES technique.
Another prediction is to control the magnitude of the defects and study if the changes in thermodynamic and transport properties due to the change in the magnitude of the pseudo-gap calculated here.
An easily testable experimental proposal  is the measurements of the slow time-dependence of the magnetic correlations and the electronic properties to which they couple. This may be done by pump-probe techniques \cite{orenstein} or low frequency microwave absorption measurements  or through the techniques of spin-echoes. Another direct prediction for experiment is unfortunately difficult to carry out. This is that the line-shape in Magnetic Bragg scattering should have a large anomalous contribution of Lorentzian squared form, given in Eq. (\ref{Sq}). Such a structure factor is observed in random-field Antiferromagnets \cite{birgeneau}. These are difficult in the case of $q=0$ transitions such as loop order, because the magnetic scattering, which is a small fraction of the usual lattice Bragg scattering, must occur atop some of the lattice Bragg peaks for magnetic patterns because there is no change in the translational symmetry. 

There is no doubt through STM measurements that there are domains formed below $T^*$ and of the size assumed and with inversion breaking. But that technique is insensitive to the time-reversal breaking feature of loop order revealed by polarized neutron scattering, which averages over domains. Measuring time-reversal breaking on the scale of domain size does not appear possible by present day techniques but may be possible in the future.

This work tries to fill the principal lacunae in a theory of the Cuprates based on three guiding principles: (1) A three orbital model is essential \cite {vsa} in the metallic state induced by doping charge transfer insulators \cite{zsa}. (2) Properties of the Strange Metal phase including its instability to Superconductivity are governed by a quantum-critical fluctuations \cite{mfl} with $\omega/T$ scaling and negligible momentum dependence \cite{ASV}. (3) The quantum-critical fluctuations are the fluctuations of a time-reversal and inversion breaking phase \cite{av} which occupies the region of the phase diagram below $T^*(x)$. 
For a theory based on these guiding principles to be credible,  all universal properties of the metallic and superconducting state of all the cuprates must follow from it and it should have verifiable and unique predictions. In the present paper, the idea that forward scattering among domains of the loop order phase generates the observed pseudo-gap is developed. This idea also has applications to other situations in where domains of $q =0$ order form through a phase transition.  \\

{\it Acknowledgements}: I have benefitted from discussions with A. Kaminski, Mike Norman,  Z.X. Shen, Inna Vishik and X.J. Zhou on results from ARPES, with Philippe Bourges, Martin Greven, Yuan Li and Y.Sidis on neutron scattering, with Seamus Davis, Abhay Pasupathy and Akash Pushp on STM results, with C. Berthier, W.P. Halperin, M.-H. Jullien,  D. MacLaughlin and A. Mounce  on NMR experiments, and with Dirk van der Marel on optical conductivity experiments.  General discussions on the matters connected with this work with Elhu Abrahams, Vivek Aji, Leon Balents, Thierry Giamarchi, Lev Gor'kov, Yan He, Steve Kivelson, Patrick Lee, Arkady Shekhter, P. Woelfle, and Lijun Zhu are gratefully acknowledged. Gian Guzman-Verri and Shaolong Chen kindly helped me with the figures. The work reported here is partially supported by the National Science foundation under grant NSF-DMR-1206298. \\


\begin{thebibliography}{99}
\bibitem{footnote}
 In the customary Ising-type model, the "spin S" at any site is fixed and the order is only in the orientation. The numerator in (\ref{Sq}) is then just $S^2$. In the class of problems considered here (which is common to order of any kind in a purely itinerant fermion system), the collective mode of interacting itinerant fermions produces a discrete order parameter which has both a variable amplitude and directions. The "$<S^2>$" coming from the amplitude is then temperature dependent. Neglecting a region of critical fluctuations in which a collective $<S^2>$ may be non-zero, the numerator
in (\ref{Sq}) is then the square of the order parameter.
\end{thebibliography}
\end{document}